\documentclass[preprint,12pt,sort&compress]{elsarticle}

\usepackage{amssymb}

\usepackage{lineno}

\usepackage{algorithm,rotating}
\usepackage{algpseudocode}
\usepackage{amsmath}
\usepackage{array, amsfonts}
\usepackage{amsthm}
\usepackage{enumitem}
\usepackage{mathrsfs}
\usepackage{float,comment}
\usepackage{mathtools}
\usepackage{todonotes}
\usepackage{hyperref}
\usepackage{listings}
\usepackage{bm}
\usepackage[utf8]{inputenc} 
\usepackage[T1]{fontenc}    
\usepackage{todonotes}

\usepackage{soul}

\usepackage{color}

\usepackage{listings}

\journal{Information Sciences}

\theoremstyle{definition}

\theoremstyle{remark}

\theoremstyle{remark}

\theoremstyle{definition}

\theoremstyle{remark}

\theoremstyle{remark}

\begin{document}

\begin{frontmatter}



\title{Minimal Algorithmic Information Loss Methods for Dimension Reduction, Feature Selection and Network Sparsification}

\author[KCL,Turing,OxfImAl,ADL,Cross]{Hector Zenil\corref{cor1}}
\ead{hector.zenil@kcl.ac.uk}
\author[ADL,OncPat]{Narsis A. Kiani}
\author[OxfImAl,Cross,WisMad]{Alyssa Adams}
\author[OxfImAl,Unicamp,LNCC]{Felipe S. Abrah\~{a}o}
\author[Vene1]{Antonio Rueda-Toicen}
\author[Vene2]{Allan A. Zea}
\author[OxfImAl,ADL]{Luan Ozelim}
\author[KAUST]{Jesper Tegn\'er}

\affiliation[KCL]{organization={Department of Biomedical Computing, Deparment of Digital Twins, School of Biomedical Engineering and Imaging Sciences},
	addressline={King's Institute for AI, King's College London},
	country={U.K.}}


\affiliation[Turing]{organization={The Alan Turing Institute},
	addressline={British Library},
	country={U.K.}}
	
\affiliation[OxfImAl]{organization={Oxford Immune Algorithmics, Oxford University Innovation and London Institute for Healthcare Engineering},
	country={U.K.}}

\affiliation[ADL]{organization={Algorithmic Dynamics Lab and Department of Oncology-Pathology, Center of Molecular Medicine},
	addressline={Karolinska Institutet},
	country={Sweden}}

\affiliation[Cross]{organization={Cross Labs},
	addressline={Cross Compass},
	country={Japan}}

\affiliation[OncPat]{organization={Karolinska Institutet},
	addressline={Department of Oncology-Pathology, Center of Molecular Medicine},
	city={Stockholm},
	country={Sweden}}

\affiliation[WisMad]{organization={University of Wisconsin-Madison},
	addressline={John W. and Jeanne M. Rowe Center for Research in Virology, Morgridge Institute for Research},
	country={USA}}

\affiliation[Unicamp]{organization={University of Campinas (UNICAMP)},
	addressline={Centre for Logic, Epistemology and the History of Science (CLE)},
	country={Brazil}}

\affiliation[LNCC]{organization={National Laboratory for Scientific Computing (LNCC)},
	addressline={Data Extreme Lab},
	country={Brazil}}

\affiliation[Vene1]{organization={Universidad Central de Venezuela},
		addressline={Instituto Nacional de Bioingenier\'{i}a},
	country={Venezuela}}

\affiliation[Vene2]{organization={Universidad Central de Venezuela},
	addressline={Department of Mathematics},
	country={Venezuela}}

\affiliation[KAUST]{organization={King Abdullah University of Science and Technology (KAUST)},
	addressline={Biological and Environmental, Science and Engineering Division},
	country={Kingdom of Saudi Arabia}\newpage}


%


%

\cortext[cor1]{Corresponding author }

\begin{abstract}
We present a novel, domain-agnostic, model-independent, unsupervised, and universally applicable Machine Learning approach for dimensionality reduction based on the principles of algorithmic complexity. Specifically, but without loss of generality, we focus on addressing the challenge of reducing certain dimensionality aspects, such as the number of edges in a network, while retaining essential features of interest. These features include preserving crucial network properties like degree distribution, clustering coefficient, edge betweenness, and degree and eigenvector centralities but can also go beyond edges to nodes and weights for network pruning and trimming. Our approach outperforms classical statistical Machine Learning techniques and state-of-the-art dimensionality reduction algorithms by preserving a greater number of data features that statistical algorithms would miss, particularly nonlinear patterns stemming from deterministic recursive processes that may look statistically random but are not. Moreover, previous approaches heavily rely on a priori feature selection, which requires constant supervision. Our findings demonstrate the effectiveness of the algorithms in overcoming some of these limitations while maintaining a time-efficient computational profile. Our approach not only matches, but also exceeds, the performance of established and state-of-the-art dimensionality reduction algorithms. We extend the applicability of our method to lossy compression tasks involving images and any multi-dimensional data. This highlights the versatility and broad utility of the approach in multiple domains.

\end{abstract}




\end{frontmatter}



\section{Introduction}

The study of large and complex datasets such as networks, has emerged as one of the central challenges in most areas of complex systems and Machine Learning, cellular and molecular networks in biology being one of the prime examples. The analysis of large-scale networks poses significant computational and analytical challenges. 

Network sparsification methods have become indispensable tools\cite{aho,batson} for addressing these challenges by reducing the complexity of networks\cite{boehmke,chew} while preserving essential structural~\cite{cunningham} or functional properties \cite{fodor,Liu2018}. Data reduction techniques extend beyond network data. Methods for image sparsification also play an essential role in managing and analysing large datasets. In essence, data reduction, in a broader sense, involves transforming various types of numerical, textual, or visual digital information into more compact representations while retaining specific properties of significance. 

This overarching concept of data reduction encompasses a wide range of methods, including network sparsification, image compression, and simplifying cellular automata rules. The common thread among these methods is the pursuit of more efficient data representation, enabling streamlined analysis and storage across various fields.

The primary objective of data dimensionality reduction in Machine Learning is twofold, much like two sides of the same coin. One facet emphasizes minimizing information loss, ensuring that valuable insights and nuances are preserved as data is transformed into a more manageable form. The other aspect underscores the importance of preserving the most `meaningful' features that uniquely characterise an object, a concept commonly referred to as feature selection. This dual objective serves as a fundamental principle in data reduction, often constituting a trade-off to be tackled in the data analysis phase.
Traditionally, identifying these meaningful features is based on a subjective criterion that is user-centric. Specific analytical goals and human experience guide the selection processes. 
For example, linear algebraic (e.g., matrix analysis) and statistically-based dimensionality reduction techniques attempt to minimise statistical information loss under certain algebraic (interpreted as signal and noise) conditions, consequently maximizing the statistical mutual information between the desired information and the dimensionally-reduced output. 

The primary objective of dimensionality reduction in network analysis is to approximate a complex network with a sparser representation. Numerous methods for graph sparsification, a form of graph summarisation, have been documented in the literature \cite{Liu2018}. Some earlier approaches, such as Chew's \cite{chew}, relied on criteria like the shortest path distance between pairs of vertices for network sparsification. Others, like Benczur and Karger \cite{bencz}, utilized cut problems as a basis for sparsification. Techniques such as spectral similarity of graph Laplacians have been employed by Spielman et al. \cite{spielman}.
When it comes to network dimensionality reduction, one must select criteria for preserving graph-theoretic properties like graph distance, clustering coefficient, or degree distribution, or combinations thereof. However, it's important to acknowledge that no finitely computable approach can comprehensively identify all possible features of interest in a dataset.
This includes recursively enumerable features that the set of all Turing machines can characterise concurrently \cite{zenildata}. Consequently, data analysts often face the daunting task of making arbitrary choices regarding which features to prioritize, as exemplified in previous research \cite{zkgraph, graphreview}.

This paper's main goal is to develop an unsupervised method that is theoretically sound and optimal in distinguishing noise from data, efficient, and independent from traditional statistical Machine Learning approaches based on classical information theory. Furthermore, the methods introduced are agnostic in selecting features of interest, encompassing all estimated computably enumerable parts within a given dataset earmarked for reduction.

We introduce a family of semi-computable algorithms designed to preserve computable properties, encompassing both statistical and algorithmic aspects. This approach is a generalisation of various dimension reduction techniques since statistically definable, generable, or recognizable patterns constitute a subset of those that algorithmic approaches can theoretically handle. 

Our method represents a novel approach to designing theoretically optimal lossy compression techniques, drawing inspiration from principles and estimations related to theoretical optimal lossless compression \cite{postchina,Kiani2016InferenceGeneticnetworks}. 
By employing efficient (polynomial) approximations~\cite{bdm,numerical} to algorithmic complexity~\cite{Calude2002} values from recent numerical methods of algorithmic probability, like in the works of ~\cite{d4} \cite{d5} and \cite{kolmo2d}, we demonstrate how these algorithms can preserve structural properties, performing similarly to, if not consistently outperforming, state-of-the-art algorithms in, e.g., the area of network dimension reduction.
We test our algorithms on non-trivial cases against transitive and non-linear (spectral) methods involving simple graphs where statistical regularities are even easier to conceal more intricate computable processes or embedded encodings, and thus may easily fool weaker~\cite{zkgraph}, linear and computable measures~\cite{zenildata,Abrahao2021}.

We will demonstrate how our method preserves critical network topological properties, including degree distribution, clustering coefficient, edge betweenness, degree, and eigenvector centralities compared to other methods. These fundamental properties provide essential information about the structure and functionality of the network.
Degree distribution, for instance, reveals how nodes are interconnected, helping us understand the overall network architecture and the distribution of connections. It is indispensable for identifying hubs and assessing network robustness.
The clustering coefficient quantifies the tendency of the nodes to form tightly connected clusters. This metric is invaluable for detecting network communities and assessing resilience against failures or attacks.
Edge betweenness measures the significance of edges in facilitating efficient communication between nodes. Identifying bottleneck edges is vital to understanding network flow and vulnerability.
The degree and eigenvector centralities help identify influential nodes within a network. High-degree nodes are hubs, while high-eigenvector centrality nodes are well-connected to other influential nodes. These centralities are vital for identifying key players in various types of networks.

By preserving these critical properties, our approach strives to provide more meaningful and accurate representations of complex networks \cite{graphreview,Abrahao2021a} while addressing the challenges inherent in data reduction techniques.
Central to our approach are perturbations in the form of edge deletions aimed at minimizing the loss of algorithmic information content \cite{Abrahao2021bpublished}. We have aptly named our algorithm MILS, signifying 'Minimal Information Loss Selection,' as it encapsulates our mission of preserving essential information while reducing data dimensionality.
Notably, the algorithms presented here remain agnostic to the specific method for approximating algorithmic complexity. They can be implemented using various approaches, including entropy or lossless compression. However, our description employs a technique based on \cite{maininfo,postchina}, which not only encompasses but also enhances other statistics- and entropic-based methods due to its intrinsic structural characteristics.

It is essential to notice that substituting the underlying methods to approximate algorithmic information content with alternatives such as Shannon entropy or lossless compression algorithms represents special cases within the broader algorithm based on algorithmic complexity. This comprehensive approach effectively covers all these less powerful cases, because the block decomposition method (BDM) calculation we employ subsumes entropy-based approximations \cite{bdm}.
BDM presents a novel approach to understanding and quantifying the complexity of biological networks and systems. BDM extends the power of the Coding Theorem Method (CTM) \cite{kolmo2d,numerical}, offering a closer connection to algorithmic complexity than previous methods based on statistical regularities. This method effectively decomposes complex objects into smaller components using a set of short computer programs. These components are then sequenced to reconstruct the original object, allowing for efficient estimations of algorithmic complexity. BDM as has been shown by our team and other researcher addresses a fundamental challenge in science: capturing the complexity of objects for classification and profiling. It bridges the gap between computable measures like Shannon entropy \cite{numerical}, which lacks invariance to object descriptions and uncomputable universal estimates of complexity. The method lies between these two, dividing data into smaller pieces to partially circumvent the uncomputability problem. 
This results in efficient estimations, albeit with some loss of accuracy. 

BDM has been applied in various fields and applications that outperform other complexity measures~\cite{Zenil2020,graphreview}, for example, to evolutionary systems~\cite{Hernandez-Orozco2018a}, artificial intelligence~\cite{Zenil2019CausalDeconv,HernandezOrozco2021AlgProbML}, and the detection of bio- and technosignature~\cite{Uthamacumaran2023arxivMSpaper}.
Significantly, the application of BDM to graphs and causal calculus has enabled the deconvolution of complex network behaviors~\cite{Zenil2018bReprogrammabilityChemicalNetworks}, identifying distinct topological generating mechanisms~\cite{Kiani2016InferenceGeneticnetworks}. This approach uses a perturbation-based causal calculus to infer model representations~\cite{Zenil2020cnat,algodyn2}, effectively tackling the challenges of causation in complex systems. We and others have demonstrated how by analysing data from discrete dynamical systems such as cellular automata and complex networks, BDM-based algorithms provide a deeper understanding of the intricate interplay of elements within these systems, revealing hidden patterns and relationships, and contributing significantly to the field of algorithmic complexity~\cite{zenilkianitegner} in biological networks~\cite{Zenil2019b,maininfo}.
Moreover, algorithmic complexity, being an asymptotically optimal measure of irreducible information content~\cite{Downey2010,vi1}, has the unique advantage of capturing algorithmically determined patterns that defy reduction to statistical patterns~\cite{Zenil2020}.

Our implementation of the MILS algorithm, rooted in algorithmic complexity, offers the added benefit of avoiding certain low-complexity deceptive phenomena. Traditional methods based solely on entropy and statistics are prone to such issues \cite{Zenil2020, zkgraph}. Significantly, this avoidance does not compromise running time efficiency or data size reduction performance for graphs, as evidenced by our results in Section~\ref{sectionResults}.  On the other hand, when implemented in ML pipelines, computational time still presents challenges.
These results demonstrate that we either match or, more often than not, surpass the performance of the best current algorithms, both for local and global graph properties. In image-compression tasks, we show that MILS presents the highest mean classification accuracy per bit of compressed images when such images are used in a particular classification task.

The approach presented in this article lies in its potential to assess the effectiveness of all reduction techniques while striving for optimal reduction by minimizing algorithmic information loss. This represents a nonlinear generalisation encompassing a broader range of techniques beyond merely preserving statistical or domain-specific algebraic properties.
While our focus in this article centers on edge deletions in graphs to describe the MILS algorithm, it's essential to emphasise that MILS can be applied to a wide array of objects and various types of constructive or destructive perturbations \cite{maininfo,Zenil2019b} on their constituent elements. For instance, MILS can seamlessly extend its applicability to node deletion, offering versatility in addressing diverse algorithmic perturbations \cite{Abrahao2021bpublished} aimed at minimizing information differences between objects.

\section{Preliminaries on complexity measures and algorithmic probability}

A comprehensive discussion of complexity measures and algorithmic probability is provided in the Supplementary Material. Meanwhile, this section introduces key preliminaries to establish the foundational concepts necessary for the main discussion. Using the principles of classical information theory, \textbf{Block Decomposition Method (BDM)} or \textbf{BDM} combines the calculation of the global Shannon Entropy rate of the object with local estimations to algorithmic complexity of smaller blocks in which the object is decomposed for which values are found in a pre-computed database of direct approximations of algorithmic probability \cite{bdm}.

The BDM relies on the following assumptions:

\begin{enumerate}
    \item In the case of small enough objects (e.g., binary strings), their \emph{algorithmic complexity} can be approximated using exhaustive search.
    \item For larger objects, breaking them into smaller parts allows for the approximation of the overall complexity by summing the complexity of individual blocks, with a correction factor to account for interactions between the blocks.
    \item For every other length, values of Shannon Entropy rates are calculated and combined with the previous values by using the same principles of information theory.
\end{enumerate}

Formally, let \( x \) be a string divided into blocks \( x_i \), with \( x = x_1 \oplus x_2 \oplus \dots \oplus x_n \), where $\oplus$ denotes a concatenation operator. The \textbf{BDM complexity} of \( x \), denoted \( \text{BDM}(x) \), is:

\[
\text{BDM}(x) = \sum_{i=1}^{n} \text{CTM}(x_i) + \log n
\]

Where:
\begin{itemize}
    \item \( \text{CTM}(x_i) \) is the algorithmic complexity approximation for block \( x_i \), derived from the Coding Theorem Method (CTM).
    \item \( \log n \) is a correction factor accounting for the interactions between the blocks.
\end{itemize}

The \textbf{Coding Theorem Method (CTM)} is a method based on the \emph{Coding Theorem} and Algorithmic Probability~\cite{vi1}, which connects classical probability to Kolmogorov complexity. The CTM maps sets of micro programs (e.g., small Turing machines) to small assembly objects for which it can empirically estimate the algorithmic probability of an object based on the following relationship~\cite{numerical}:

\[
K(s) \approx -\log P(s)
\]

Where:
\begin{itemize}
    \item \( K(s) \) is the Kolmogorov complexity of string \( s \).
    \item \( P(s) \) is the algorithmic probability of string \( s \), as defined by Solomonoff's distribution.
\end{itemize}

The CTM leverages pre-computed distributions from small Turing machines to approximate the algorithmic complexity of small strings by estimating their algorithmic probability~\cite{d5,kolmo2d}. BDM and CTM can be seen as practical approximations of algorithmic probability and Kolmogorov complexity, especially in scenarios where exact computation is impossible. The relationship can be summarised as follows:

On the one hand, \textbf{CTM} provides an approximation to \emph{algorithmic probability} \( P(s) \) by connecting the empirical frequency of a string's generation in a small Turing machine model to its \emph{Kolmogorov complexity} \( K(s) \), using the relation \( K(s) \approx -\log P(s) \).
On the other hand, \textbf{BDM} offers a method to map micro programs to their respective small pieces of an object requiring an algorithmic explanation.

\subsection{Algorithmic information theory, data compression and feature selection}

BDM, CTM and Algorithmic Probability offer a principled approach to data compression and feature selection by quantifying algorithmic complexity rather than relying on traditional statistical methods. In data compression, these methods allow for data decomposition into patches and estimating their Kolmogorov complexity, ultimately selectively removing redundant regions while preserving structurally meaningful information. This empowers compression routines which prioritize algorithmically significant structures rather than merely frequent ones, resulting in compression that minimises storage requirements while retaining essential data properties.

In feature selection, these methods provide a more fundamental measure of information content than conventional statistical metrics like correlation or variance. These methods help identify which features contain essential computational information, enabling the removal of redundant or low-complexity features without significant loss of structure. Thus, their ability to optimize data representation while retaining essential information makes them valuable tools for scientific computing, machine learning, and complex data-driven applications.

\section{Method}\label{sectionMethod}

\subsection{Calculating information difference values}\label{informationdifference}
All methods employed in this article are based on \emph{information difference} among an object's elements. Specifically, we focus on the contribution of each element to the system's overall information content, such as the impact of nodes or links in a network context. Formally, this involves assessing the impact of \emph{local} perturbations (changes made to individual elements) on the algorithmic information content of the entire object, compared to its prior state.
These methods are grounded in algorithmic/causal perturbation analysis, a key component of algorithmic information dynamics (AID) \cite{Zenil2020cnat,algodyn2}. In this contribution, our approach involves the \emph{perturbation} of the elements of the system through the removal of components, followed by measuring and ranking the effects on the algorithmic information content of the system. This dynamic approach to studying and calculating the potential changes in an object to assess the contribution of its components was initially introduced in~\cite{zenilgraph,algodyn}. 

We extend these concepts to data/network dimension reduction and lossy compression of bi-dimensional data, such as images.
For a graph $G$ with edges $e_1, \ldots, e_{|E(G)|} \in E(G)$, we consider the subset $F \subseteq E(G)$ representing single-edge perturbations. The graph $G \backslash F$ results from removing edges in $F$. The algorithmic complexity $C(G)$ (as detailed in the \emph{Supplementary Material}) allows us to define the information difference for edge deletion:
\begin{equation}\label{equationInformationdifferencefordeletion}
I(G,F) = C(G) - C(G \setminus F)
\end{equation}
which quantifies the \emph{information difference} of each edge $e_i$ in the context of edge deletion.
Notice that this difference could be defined for perturbations on the graph's vertices, but here, we will restrict our attention to perturbations on edges.
In the present article, as introduced in the next Section~\ref{sectionMILSalgo}, we focus on the empirical investigation of destructive perturbations (i.e., the edge deletion problem) to perform network reduction while minimizing information loss.
However, in the Sup. Mat. we also theoretically investigate the edge-insertion case as an analogous inverse case of edge deletion.
Future research will be fruitful for empirically correlating the present results in Section~\ref{sectionResults} (which hold for destructive perturbations) with those for constructive perturbations.

\subsection{An algorithm for minimal information loss selection}\label{sectionMILSalgo}

MILS is an unsupervised and mostly parameter-free algorithm, i.e., asymptotically independent of model or domain, as it does not need to be instructed or designed to preserve any particular property, and maximises the preservation of all computable elements that contribute to the algorithmic information content of the data.
\begin{algorithm}[hb!]
	\caption{Minimal Information Loss Selection (MILS)}\label{milsalg}
	\begin{algorithmic}[1]
		\footnotesize
		\Function{InfoRank}{$G$}
		\State $\textit{informationLoss} \gets \varnothing$
		\Statex\hspace{1.5em}{\color{gray} // for each edge $e_i$}
		\For{$e_i\in E(G)$}\label{stepInfoRankloop}
		\Statex\hspace{3em}{\color{gray} // store information contribution into {\it informationLoss}}
		\State $\textit{informationLoss} \gets \textit{informationLoss} \cup \{I(G,e_i)\}$
		\EndFor
		\State \textbf{sort} $\textit{informationLoss}$ \textbf{in increasing order}
		\Statex\hspace{1.5em}{\color{gray} // return information rank}
		\State \Return{$\textit{informationLoss}$}
		\EndFunction
		
		\vspace{2mm}
		
		\Function{MILS}{$G,N$}, $1\leq N\leq |E(G)|$
		\While{$|E(G)|>N$} \label{stepMainloopMILS}
		\Statex\hspace{3em}{\color{gray} // calculate minimal loss across all edges}
		\State $\textit{minLoss} \gets \min(\textsc{InfoRank(G)})$\label{stepCallInfoRank}
		\Statex\hspace{3em}{\color{gray} // remove all candidate edges from $G$}
		\State $\textit{G} \gets G\backslash\{e_i\in E(G):I(G,e_i)=\textit{minLoss}\}$
		\EndWhile
		\State \Return{$G$} \label{stepReturnMILS}
		\EndFunction
	\end{algorithmic}
\end{algorithm}

\begin{figure}[ht!]
	\centering
	{ \hspace{-2.5cm}\emph{MILS Tree Diagram}\\ \vspace{0.35cm} }
	{\centering
	\includegraphics[width=\linewidth]{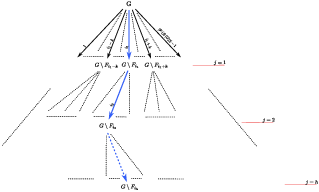}
	\caption{\label{figTreeDiagram} \emph{Tree Diagram} for Algorithm~\ref{milsalg} with the sequence of graphs with their respective subsets of edges reduced (as selected in the iterations $ j $ of [Algorithm~\ref{milsalg}, Step~\ref{stepMainloopMILS}]).
	So, one has that $ F_{ i_1 } = \textit{minLoss}_1 = \min\left( \textsc{InfoRank}\left( G \right) \right)$, $ F_{ i_2 } = \textit{minLoss}_2 =\min\left( \textsc{InfoRank}\left( G \setminus F_{ i_1 } \right) \right)$, and so on, which continues up until the final graph $ G \setminus F_{ i_h } $ is found by the algorithm.
	This process is represented by the blue path $ \left( G , G \setminus F_{ i_1 } , G \setminus F_{ i_2 } , \dots , G \setminus F_{ i_h } \right) $ from the root $ G $ to the leaf $ G \setminus F_{ i_h } $ 
	in $ h $ iterations.
	Notice that from each parent vertex/graph, there are as many children as (non-empty) subsets of (present) edges in the parent. 
	Fig.~\ref{figTreeDiagram} illustrates a speed-up both globally (as the number of iterations $ j \leq h $ is always upper bounded by $ \mathbf{O}\left( \left| E\left( G \right) \right| - N  \right) $) and locally (as from each parent, Algorithm~\ref{milsalg} does \emph{not} have to search for the best candidate across all possible subsets of edges of the parent) achieved by Algorithm~\ref{milsalg} from the decision process that this tree diagram represents.
	}
	}
\end{figure}

Let $G$ be a graph, and $ \left| E\left( G \right) \right| $ denote its number of edges. 
MILS seeks to reduce $G$ to a graph on $ N < \left| E\left( G \right) \right| $ edges so that the loss of information is minimised. 
Ideally, one would do so by calculating the information differences between $ G $ and $G\backslash S$, where $ \left| S \right| \geq 1 $, and then finding the subset $F\subset E(G)$ such that $I(G,F)\leq I(G,S)$ for all non-empty proper subsets of edges $S\subset E(G)$, repeating this task with $G \coloneqq G\backslash F$ until the target size is reached (i.e., when $|E(G)|=N$). 
This algorithmic procedure's time complexity is clearly in $O(exp)$ because of how it performs searches and deletions across all subsets of edges (see also Fig.~\ref{figTreeDiagram}), but significant improvements to this bound are still possible as shown by Algorithm~\ref{milsalg} (see also Algorithms~$ 2 $ and $ 3 $ in the Sup. Mat.).

\begin{figure}[ht!]
		\centering
		\includegraphics[width=0.8\textwidth]{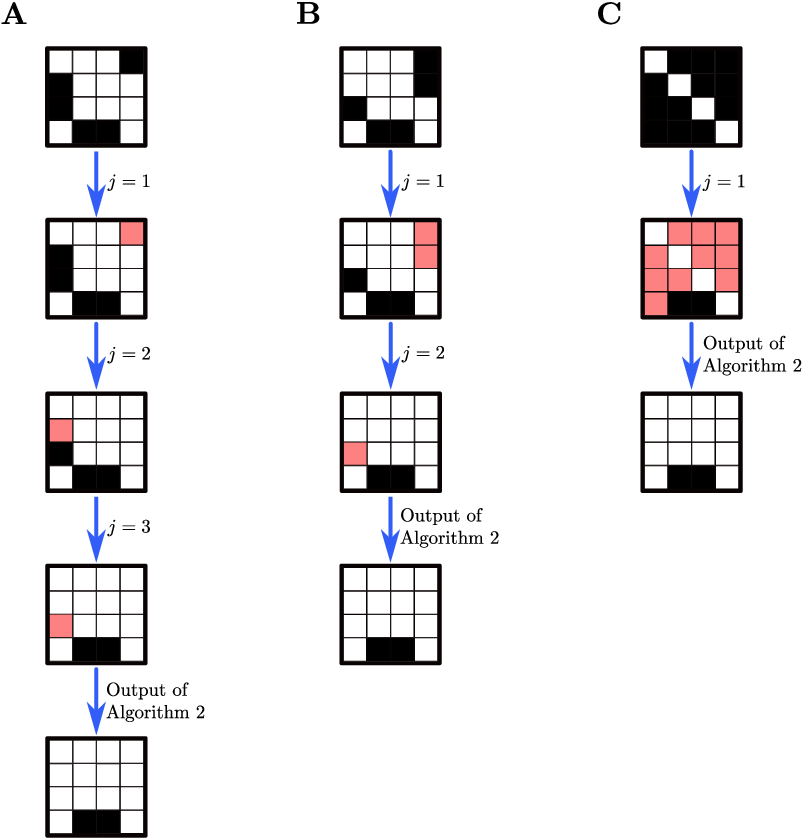}
	\caption{\label{figGridDiagram} Three examples of edge deletions occurring in each iteration $ j $ of [Algorithm~\ref{milsalg}, Step~\ref{stepMainloopMILS}], indicated by the downward blue arrows (except for the bottommost blue arrows), while the subsequent red squares indicate where the deletions occurred in the previous step. 
	The bottommost blue arrows in each Fig.~\ref{figGridDiagram}\textbf{A}, \ref{figGridDiagram}\textbf{B}, and \ref{figGridDiagram}\textbf{C} refer to [Algorithm~\ref{milsalg}, Step~\ref{stepReturnMILS}] after which the final output of the algorithm is returned, given an adjacency matrix of a graph $ G $ (such as anyone of the three topmost $ 4 \times 4 $ matrices) as input.
	The input adjacency matrix in Fig.~\ref{figGridDiagram}\textbf{A} corresponds exactly to a substring of the halting probability as in \cite{Calude2022GlimpseOmega}. 
	Mixing both redundant and incompressible components, the second row of the input matrix in Fig.~\ref{figGridDiagram}\textbf{B} is a repetition of the first row, and the remaining $ 8 $ bits equals to those of the input matrix in Fig.~\ref{figGridDiagram}\textbf{A}.
	Fig.~\ref{figGridDiagram}\textbf{C} presents one of the most redundant cases in which all oriented edges are present, that is, the input is an adjacency matrix of a complete (directed) graph.
	}
\end{figure}

A more efficient but suboptimal version of such an exponential-time MILS iterates over single elements (in the case, edges) or singletons. 
Algorithm~$ 2 $ in the Sup. Mat. performs such a sequential deletions of edges, and then removes the edge that contributes less to the information content of the graph before moving to the next step.
When $e$ is such that $ \left| I(G,e) \right| $ is minimal, where $ I(G,e) $ denotes $ I( G , \{ e \} ) $ (i.e., the information difference from deleting only a single edge), we call it a \emph{neutral information edge} \cite{maininfo} because it is an edge that contributes the least to the information content of $G$ (in particular, it minimises information loss or the introduction of spurious information into the network according to the information difference when removed from the original network).
The existence of such edges whose information is maximally neutral (or redundant) is demonstrated in Theorem~$ 3.8 $ in the Sup. Mat., where it is shown that there are infinite families of graphs for which each member has at least one maximally neutral information edge $ e_i $ (in the edge deletion case), where $ \left| I(G,e_i) \right| = \left| C(G) - C( G\backslash e_i ) \right| = \mathbf{O}( 1 ) $. 
This is the opposite case of a \emph{maximal information edge}, i.e., a single edge whose deletion changes the most the information content of the entire $ G $, case which refers to incompressible graphs (see Section~\ref{sectionTheory} and Fig.~\ref{figGridDiagram} discussed below).
Thus, although suboptimal in general, our application of more efficient versions of MILS based on single edges is capable of grasping the full range between minimal and maximal information contribution to the whole network.
See Section~\ref{sectionTheory} for the information difference error of our efficient algorithms in comparison to the optimal exponential-time MILS.

The pseudocode in Algorithm~$ 2 $ in the Sup. Mat. assumes that there is a unique such $e$, which may not necessarily be the case in general. 
Algorithm~\ref{milsalg} solves this problem by performing simultaneous deletions on all edges with an information contribution given by those in $\textit{minLoss}$.
To this end, Algorithm~\ref{milsalg} introduces \textsc{InfoRank}, a method that produces a ranking of $e_1,\dots,e_{|E(G)|}$ from least informative to most informative edge, i.e., a list of edges sorted in increasing order by their information contribution to $G$.
This ranking facilitates the search for the (most) neutral elements of the system (see Sections~$ 2.8 $ and~$ 3.3 $ in the Sup. Mat.), which in turn helps MILS preserve the components that maximise the information content of the resulting object with a speed-up in running time compared to the ideal exponential-time version of MILS---a speed-up achieve even in case neutral elements are not unique. 
See also Section~\ref{sectionTheory} for a theoretical discussion.

Fig.~\ref{figGridDiagram} presents three illustrative examples of Algorithm~\ref{milsalg} returning the same graph (with $ N \leq 2 $ present edges) as output for three distinct inputs displayed in Fig.~\ref{figGridDiagram}\textbf{A}, Fig.~\ref{figGridDiagram}\textbf{B}, and Fig.~\ref{figGridDiagram}\textbf{C}, respectively.
The topmost matrices in columns \textbf{A}, \textbf{B}, and \textbf{C} display the adjacency matrices of distinct directed graphs (without self-loops), all of them with four vertices, that are given as inputs.
Each of these graphs are subjected to the edge deletions that Algorithm~\ref{milsalg} performs so as to generate its final output.
As the adjacency matrix given as input becomes less random, one should expect that the number of iterations can be reduced.
Indeed, this is what happens in Fig.~\ref{figGridDiagram}\textbf{A}, ~\ref{figGridDiagram}\textbf{B} and~\ref{figGridDiagram}\textbf{C}, respectively:
in example \textbf{A}, the input matrix is (algorithmically) random, i.e., incompressible, hence no edge in the (ordered) list given by \textsc{InfoRank} will be as neutral as the other ones, and \textit{minLoss} will contain only \emph{one} edge at the time;
in example \textbf{B}, the edge in the first row is a neutral information edge as much as the edge in the second row, and therefore both edges will be deleted in only one iteration step;
in example \textbf{C}, except for the last two edges in the last row, every edge will be considered neutral in \textsc{InfoRank}.
Further discussion on Fig.~\ref{figGridDiagram} can be found in Section~$ 2.8 $ in the Sup. Mat..

The \emph{tree diagram} in Fig.~\ref{figTreeDiagram} illustrates the decision process involved in each iteration of Algorithm~\ref{milsalg} that results in the next size-reduced graph, speeding up this process both \emph{locally} and \emph{globally} across the tree.
Each child vertex represents a graph resulting from the deletion of one or more edges in the parent vertex (i.e., a graph before the destructive perturbations on the edges). 
In each descending step starting from the root at the top, as the height of the tree downward increases, a subset of (present) edges is deleted from the parent graph at the lower level.
Notice that from each parent vertex, there can be as many child vertices as the size of the power set of the edges (excluding the empty set) in the graph represented by the parent vertex.
The child vertex/graph chosen by Algorithm~\ref{milsalg} is the one whose neutral information edges (according to \textsc{InfoRank}) were deleted from the parent vertex/graph.
In this manner, the iteration process in [Algorithm~\ref{milsalg}, Step~\ref{stepMainloopMILS}] is represented by the rooted path highlighted in the color blue in the Fig.~\ref{figTreeDiagram}.
A subset of edges that has neutral information with respect to the parent graph is deleted from the parent graph so as to generate the child graph in this path.
Thus, \textsc{InfoRank} allows Algorithm~\ref{milsalg} to select a child graph that minimises the information difference with respect to the parent graph, not having to search over all the possible subsets of present edges.
This single-thread process continues until the leaf vertex, i.e., the graph with (at most) $ N $ edges is achieved.

This allows a \emph{local} speed-up between each parent graph and its children in Fig.~\ref{figTreeDiagram}, in particular by replacing a search over an exponential-size set of subsets with a search over a list (of edges), therefore quadratic with the number of vertices/nodes in the root graph.
It also allows a \emph{global} speed-up between the root vertex and the final output of Algorithm~\ref{milsalg}, particularly by replacing a search over all possible rooted paths with a recursive iteration in [Algorithm~\ref{milsalg}, Step~\ref{stepMainloopMILS}] whose maximum number of iterations is always upper bounded by $ h \leq \left| E\left( G \right) \right| - N + 1 $, therefore quadratic with the number of vertices in the root graph.
This makes the height of the tree always much shorter in comparison to its width.

It is also important to notice that the decision tree diagram (i.e., the one for the minimisation of the information difference between the parent and the child vertex) in Fig.~\ref{figTreeDiagram} can be subsumed into a \emph{probability tree diagram}.
As shown in Section~$ 2.8.1 $ 
in the Sup. Mat., each path from the root to a leaf has an associate joint probability of a sequence of independent events, and each event corresponds to a destructive perturbation on a subset of edges.
The rooted path highlighted in the color blue in the Fig.~\ref{figTreeDiagram} illustrates the path to the final leaf vertex (i.e., the vertex representing the outcome of Algorithm~\ref{milsalg}) that \emph{maximises} the (algorithmic) probability of edge deletions, while each edge in this path corresponds to an iteration in [Algorithm~\ref{milsalg}, Step~\ref{stepMainloopMILS}].
As demonstrated in Section~$ 2.8.1 $ 
in the Sup. Mat., which in turn is a result that evinces the generality of our methods, the MILS tree diagram in Fig.~\ref{figTreeDiagram} shows that a decision process of minimisation of information differences implies a maximization of (algorithmic) probabilities both locally for each arrow and globally for each path from the root to a leaf. 
This holds theoretically independent of the choice of the computation model, programming language, or any arbitrary (computable) probability (semi-)measure that one aims at maximizing (whether or not one assumes independence between parent vertices and subsequent child events in such a measure).
See Section~$ 2.8.1 $ in the Sup. Mat. for a thorough explanation of the tree diagram and its mathematical properties.

As described in \cite{maininfo} and discussed in the Sup. Mat., one can easily modify Algorithm~\ref{milsalg} to deal with multiple edge deletions at once, as in $ I(G,F) $, but possibly at the expense of much more computational resources. 
In addition, note that these algorithms may be applied, \textit{mutatis mutandis},  to nodes or to any data element of an object or dataset, e.g. a pixel of an image, or a row or column in a spreadsheet, which constitutes an exciting application of our current methods as shown in Section~\ref{sectionMILSonimages}.

\section{Theory}\label{sectionTheory}

In this section, we introduce the theoretical results that support the methods employed in Section~\ref{sectionResults}.

As presented in Section~\ref{informationdifference} and formalized in \cite{Abrahao2021bpublished,Zenil2020cnat,maininfo,zenilkianitegner}, we have that edge deletions can be seen as transformations/perturbations that a particular network is being subjected to.
This way, it is immediate to ask about the maximum overall impact of these transformations on the algorithmic information content of the network.
Indeed, in Section~$ 3.1 $ in the Sup. Mat. we show that the theoretically optimal information difference (i.e., the information difference calculated by the exact theoretical algorithmic complexity values) produced by edge deletions are bounded by terms that depend on the number of single-edge perturbations, the network size and/or graph sparsity (or edge density).
In the general case, Theorem~$ 3.1 $ in the Sup. Mat. demonstrates that algorithms for approximating $ I(G,F) $ ideally calculate values that fall into the range of values upper- and lower-bounded by terms in a \emph{logarithmic} order of the \emph{network size} $ N = \left| V(G) \right| $ and in a \emph{linear} order of the \emph{number} $ \left| F \right| $ \emph{of single-edge perturbations}, except for other additive terms that are strongly dominated by the former two terms, respectively.
As shown in Corollary~$ 3.2 $ in the Sup. Mat., this behavior remains the same even in case the graph is (algorithmically) random, i.e. \emph{incompressible}---as it is the case of most sufficiently large Erd\H{o}s-R\'enyi (ER) \emph{random graphs} with edge probability $ 1/2 $---, so that even in the case each edge theoretically carries the maximum information it can, or carries the least amount of redundancies (that is, each edge contributes as much as the others to the overall irreducible information content of the graph), those upper and lower bounds hold.

As also shown in Sup. Mat., by replacing network size with edge density (i.e., the size of $ \left| E(G) \right| $ in comparison to $ \frac{ N^2 -N }{2} $), we demonstrate even tighter lower and upper bounds for multiple edge deletions that are in a logarithmic order of the edge density and, improving on the general case to the above, in direct proportion to the number $ \left| F \right| $ of single-edge perturbations.
Thus, our results demonstrate that these limits for the algorithmic information loss (or gain) from edge perturbations have a linearly dominant dependence on the number of single-edge perturbations, which is denoted by $ \left| F \right| $, except for a multiplicative logarithmic term depending either on the network size or on the edge density.
In turn, these strongly dominate all the other reminiscent additive terms.

In addition, as expected in an alternative approach for estimating those lower and upper bounds but based on a straightforward and direct encoding of the subsets of edges to be perturbed---therefore, \emph{more simplistic} than the proof techniques employed in this work---, we demonstrate that one can always retrieve other lower and upper bounds which (although not as tight as the ones we obtained) do only depended on the network size, but now in a \emph{quadratic} order. 
See Corollary~$ 3.3 $ in the Sup. Mat..

Moreover, as should be expected for only \emph{one} single-edge perturbation (i.e., when $ \left| F \right| = 1 $), we demonstrate that one can also derive that these upper and lower bounds are still tight in general (see Corollary $ 3.4 $ in the Sup. Mat.), but in this case only depending on a logarithmic order of the network size.

These theoretical results pave the way for understanding why MILS algorithms are optimal with respect to their chosen edge-deletion scheme, respectively, once one fixes the arbitrarily chosen method capable of approximating the algorithmic complexity values in the asymptotic limit.
That is, the algorithms return smaller networks whose respective edge count reductions are done so that the information difference between the original network and the smaller one is minimised according to those optimal upper and lower bounds described in the above paragraphs.
This holds not only for the ideal EXPTIME version of MILS mentioned in Section~\ref{sectionMILSalgo}, whose iterations cover the power set of the set $ E( G ) $ of present edges, but our results show that switching to a summarisation procedure based on successive single-edge deletions (like in Algorithm~\ref{milsalg}) is sound.
In particular, as discussed in this section, the latter case is grounded on the fact we proved that any maximum error does not dependent on the particular ideal subset of edges to be deleted, but only on the sheer size of this subset (once the worst-case logarithmic contribution of the network size becomes fixed).
This means that in the theoretical limit with unbounded resources, Algorithm~\ref{milsalg} can never misestimate the minimisation of information difference by an \emph{error} larger than a linear factor of the number of single-edge perturbations that should otherwise be ideally deleted.
Even if one compares the Algorithm~\ref{milsalg}'s results with the ones obtained from the best deletion scheme (i.e., the one in EXPTIME), the maximum information difference error (when choosing a scheme distinct from the ideal exponential-time MILS) increases linearly with a logarithmic term of the network size as the number of single-edge perturbations increases.
Thus, this error is far from the quadratic dependence on the network size (see Corollary~$ 3.3 $ in the Sup. Mat.) that one might expect from a more simplistic method of encoding subsets of edges, as discussed to the above.

Theorem~$ 3.11 $ in the Sup. Mat. demonstrates that Algorithm~\ref{milsalg} not only returns a unique output for each input (which holds even in the case of more than one neutral element at each iteration, as discussed in Section~\ref{sectionMILSalgo}) but also has: a worst-case time complexity with a \emph{cubic} dependence on \emph{input} $ E( G ) $, at the same time that it scales \emph{linearly} with the worst-case time complexity of the \emph{chosen method} for approximating the algorithmic complexity values of any edge.

Thus, our methods employed in Section~\ref{sectionResults} take advantage of a trade-off between the information difference \emph{error} and the running time \emph{speed-up} in which:
the information difference can only diverge from the ideal one in the EXPTIME MILS algorithm by a linear order of the number of single-edge perturbations, divergence which does not depend on the ideal subset of edges that could have been otherwise deleted in the EXPTIME version of MILS;
while the speed-up improves from an exponential to a cubic worst-case time complexity.
This speed-up is illustrated by the tree diagram in Fig.~\ref{figTreeDiagram}.
Instead of searching along all possible paths, which would demand exponential running time, each iteration corresponds to an one-step increase in the height of the tree from the root, and in each step the algorithm runs the \textsc{InfoRank} procedure in order to select the edge deletions to be made.

\subsection{How and Why MILS works}

The MILS process is a greedy algorithm that begins with patch decomposition, where the input data--an image, matrix, or graph--is divided into smaller regions or patches. These patches have their algorithmic complexity estimated using BDM, which quantifies how difficult these patches are to generate as the result of a short-length program (or many of them, according to Algorithmic Probability). During the deletion phase, MILS applies perturbations such as edge deletions in graphs or bit-flipping in images while maximizing preservation of the size of the shortest programs that generate the data. Instead of relying on conventional downsampling techniques, which often distort structural information, MILS strategically removes elements in a way that minimally affects BDM (its algorithmic complexity estimation), thereby attempting to preserve the original dataset's causal, recursive or algorithmic structure. If after removing an element the complexity of the 2D array does not change, with high probability the removed element is not relevant to the generative process itself and can therefore be safely removed as compared to other more relevant.

Then coarse-graining is performed by ensuring that the algorithmic complexity of each patch remains nearly constant even after deletion, meaning that the essential features of the data are retained despite the data dimensionality reduction. The remaining patches are then stitched together in a way that maintains their original relationships. If a region of the dataset is highly compressible and exhibits low BDM, it can tolerate more deletions without significantly affecting its algorithmic information content connected to its causal content.
Conversely, highly complex regions with high BDM are preserved more carefully, as they contain intricate patterns that are difficult to reconstruct from limited data.

Ultimately, MILS does not remove data arbitrarily but rather follows an algorithmic-complexity-preservation strategy, selecting deletions that minimise changes in the generative complexity of the data. By treating the dataset as a set of algorithmic programs and ensuring that deletions do not substantially alter the shortest program needed for reconstruction, MILS effectively reduces dimensionality while maintaining structural integrity. This makes it highly valuable for applications in data compression, robust machine learning, and high-dimensional data analysis, where retaining essential information is critical.

\section{Results}\label{sectionResults}

\subsection{Experimental Validation and Benchmarking of MILS for Dimensionality Reduction in Machine Learning Pipelines}

As indicated, MILS method offers a novel approach by leveraging principles of algorithmic complexity to achieve superior compression with minimal degradation of accuracy. MILS is particularly relevant in quantisation scenarios where preserving information per bit is crucial, such as low-power edge computing, compressed neural networks, and large-scale retrieval-augmented systems (RAGs). By incorporating MILS into ML pipelines, models can maintain higher accuracy per bit of compressed data, making it an attractive alternative to conventional dimension reduction methods.

To evaluate the efficiency of MILS, we conduct a benchmark study focusing on classification accuracy performance per bit of compressed image. This metric provides insight into how well different dimension reduction methods retain information relative to their storage and computational cost. The MILS method is compared against several established techniques across different quantisation levels to determine its effectiveness in preserving predictive performance while minimizing information loss.

We utilize the MNIST dataset from the `sklearn` library as the benchmark dataset for this study. It consists of 1797 8x8 images of digits, encompassing 17 features in each image (17 grayscale tones). There are about 180 samples per digit class in this dataset and it is suitable for quick benchmarking due to its relatively small size. Given its widespread use in evaluating ML techniques, MNIST serves as an ideal benchmark to test the efficacy of MILS in a controlled setting.

Three levels of quantisation are considered for our analysis and were obtained after using the `gfloat' Python package:

\begin{itemize}
    \item 1-bit binary representation (used by MILS)
    \item 4-bit floating point representation (OCP E2M1)
    \item 8-bit floating point representation (OCP E4M3)
    \item 16-bit floating point representation (Brain Floating Point bfloat16)
\end{itemize}

Besides, MILS is compared against three classes of dimension reduction techniques:

\begin{itemize}
    \item Matrix decomposition-based methods: Principal Component Analysis (PCA) and Non-Negative Matrix Factorisation (NMF).
    \item Random projection-based methods: Sparse Random Projection (SPJ) and Gaussian Random Projection (GRJ).
    \item Spectral methods: Kernel PCA with radial basis function (KPCAR) and polynomial kernels (KPCAP). The link of Kernel PCA with spectral methods was previously established in \cite{Bengio}.
\end{itemize}

To properly compare compression techniques of images, we need to consider their reconstructed counterparts to assess whether information loss has occurred. Thus, rather than using the compressed representations directly, we analyze the reconstructed images to determine if the compression workflow led to a degradation of relevant information. Thus, the following Machine Learning pipeline was implemented to perform the benchmarks:

\begin{itemize}
    \item Create an image compression step, where different models were used to compress the original 8x8 images. The output of this step is also a 8x8 image (the one coming from reconstruction).
    \item From the compressed-reconstructed version of the 8x8 image, train a Support Vector Classifier to classify the dataset according to the MNIST classes (digits from 0 to 9). Since we will perform model selection, a 5-fold StratifiedShuffleSplit cross-validation (CV) is performed with test size equal to 20\% of input samples, ensuring that class balance is retained during training and that samples are shuffled. For simplicity, no hyperparameter tuning will be performed, thus a validation dataset is not needed. For the CV results assessment, the mean balanced accuracy metric was chosen (the average of recall obtained on each class).
\end{itemize}

This Pipeline was repeated for varying compression rates, hereby governed by the number of components used in traditional techniques or the number of edges kept in the MILS workflow. We chose the following values for these: 1,2,4,8,16,32 and 64, being the latter equivalent to using the full uncompressed image as input.

Besides the pyMILS implementation described in the Supplementary Material, we used another implementation which considers Algorithm 2 (where each edge is visited to check for candidate deletion cases). This MILS implementation is shown in Listing 1.

\begin{lstlisting}[language=Python, caption= MILS core function]
bdm = BDM(ndim=2,partition=PartitionCorrelated)
pe = PerturbationExperiment(bdm, metric='bdm')

def MILS(X,N):
    X_red = X.copy()
    idx = np.argwhere(X_red)
    while idx.shape[0]>N:    
        pe.set_data(X_red)
        delta_bdm = pe.run(idx)
        b = idx[np.argmax(-np.abs(delta_bdm))]
        X_red[*b] = 0
        idx = np.argwhere(X_red)
    return X_red
\end{lstlisting}

Such implementation relies on the `pyBDM' Python package and its implementation of a Perturbation Experiment (which changes elements in a 2D array and calculates the resulting change in BDM). By using this function, we can define the compressor-reconstruction step for the MILS algorithm as shown in Listing 2.

\begin{lstlisting}[language=Python, caption= MILS step]
def MILS_step(N,input_series):
    s1,s2 = input_series.shape
    mils_v = Parallel(n_jobs=32)(delayed(MILS)(*x) for x in [[z,N] for z in input_series.reshape(s1,8,8)])
    return np.array(mils_v).reshape(s1,s2)
\end{lstlisting}

In short, MILS\_step performs the deletion of edges for the whole input dataset passed to the function. Several input samples are processed in parallel to enhance the speed of the transformation. It is important to notice that we consider that the compressed and reconstructed versions of an image after MILS are the same, following the rationale that changes are carried out in place. This compression-reconstruction step is then incorporated to the full Pipeline as presented in Listing 3.

\begin{lstlisting}[language=Python, caption= MILS Pipeline]
start = time.monotonic()
perf_MILS=[]
for nel in [2**k for k in range(7)]:
    transformerMILS = FunctionTransformer(partial(MILS_step,nel))
    sss = StratifiedShuffleSplit(n_splits=5, test_size=0.2, random_state=0)
    pipelineMILS = Pipeline([('transformer', transformerMILS),
                         ('classifier', SVC(gamma='auto'))])
    scr = cross_val_score(pipelineMILS, (X>7).astype(int), y, n_jobs=-1, cv = sss,scoring='balanced_accuracy')
    perf_MILS.append([nel,scr])
end = time.monotonic()
general_time.append(["MILS",end-start])
\end{lstlisting}

It can be seen that prior to the application of MILS, the data is binarized by simply attributing 1 to feature values greater than 7 and 0 otherwise. By running the Pipeline in Listing 3, the performance of MILS can be assessed. To illustrate how other methods were implemented, Listing 4 presents the equivalent Pipeline for PCA reduction technique.

\begin{lstlisting}[language=Python, caption= PCA Pipeline]
def PCA_step(N,input_series):
    mdl = Pipeline([('transformer', StandardScaler()),('pca', PCA(n_components=N))])
    mdl = mdl.fit(input_series)
    mdl_outp = mdl.named_steps["pca"].inverse_transform(mdl.transform(input_series))
    return mdl_outp

perf_PCAs_quant=[]
for fmt,tm in zip([format_info_ocp_e2m1,format_info_ocp_e4m3,format_info_bfloat16],["PCA_e2m1","PCA_e4m3","PCA_bfloat16"]):
    start = time.monotonic()
    perf_PCA_q=[]
    for nel in [2**k for k in range(7)]:
        sss = StratifiedShuffleSplit(n_splits=5, test_size=0.2, random_state=0)
        transformerPCA = FunctionTransformer(partial(PCA_step,nel))
        pipelinePCA = Pipeline([('transformer', transformerPCA),
                                ('quantizer',FunctionTransformer( lambda x: round_ndarray(fmt, x,sat=True))),
                             ('classifier', SVC(gamma='auto'))])
        scrPCA = cross_val_score(pipelinePCA, round_ndarray(fmt, X,sat=True), y, n_jobs=-1, cv = sss,scoring='balanced_accuracy')
        perf_PCA_q.append([nel,scrPCA])
    perf_PCAs_quant.append(perf_PCA_q)
    end = time.monotonic()
    general_time.append([tm,end-start])
\end{lstlisting}

Listing 4 presents the implementation of a PCA\_step, similar to MILS\_step, where the compression and reconstruction is carried out. Differently from the case of MILS, where only a binary quantisation is considered, for all the other methods the 4-bit,8-bit and 16-bit quantisations are used. Before fitting each model, no additional preprocessing techniques were applied except for the PCA case, where a StandardScaler() (subtracts the mean and divides by the standard deviation of each feature) was applied. Also, all the simulations considered the `scikit-learn' implementation of the algorithms. 

\begin{figure}[ht!]
	\centering
	\includegraphics[width=0.95\textwidth]{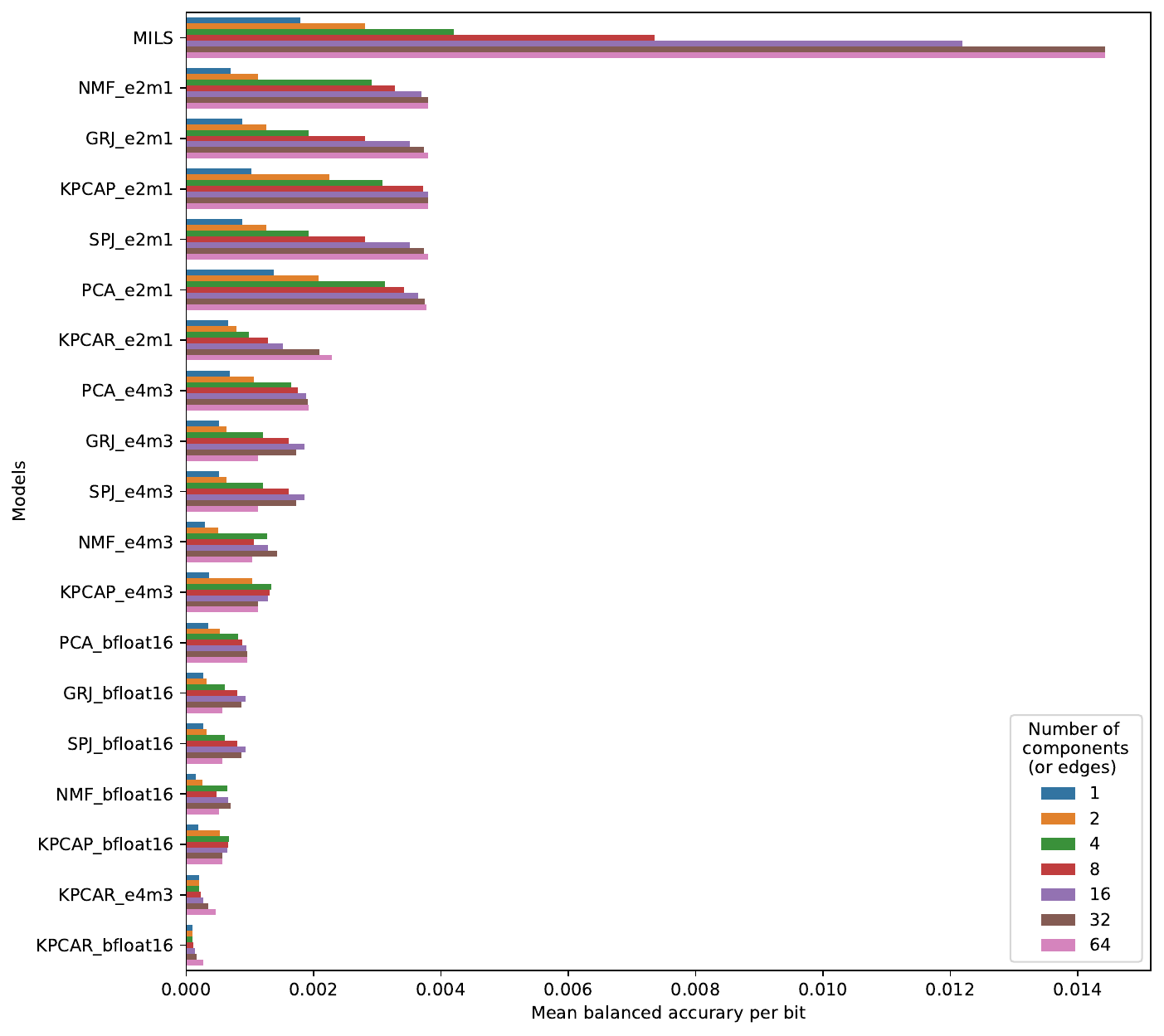}
	\caption{\label{bench} Performance assessment (mean accuracy per bit of compressed image) of different compression algorithms, including MILS and other traditional approaches. MILS substantially outperforms all the considered algorithms due to its highly compressed binary setting.}
\end{figure}

Figure \ref{bench} presents the performance assessment results, which demonstrate that MILS consistently delivers higher accuracy per bit across all quantisation scenarios in the MNIST classification task. The mean balanced accuracy per bit is calculated as the mean balanced accuracy of the 5-fold CV divided by $64 \times q$, where $q$ is the quantization level of the compressed-reconstructed figure.

Notably, as the number of components/edges is reduced, the performance gap between MILS and competing methods narrows, but MILS maintains a lead in all tested cases and quantisation levels. We attribute the superiority of MILS to its ability to optimally preserve essential structural and algorithmic complexity features of the data, leading to better generalisation in classification tasks. This is done while keeping the memory footprint to a minimum, since only binary values are used.

\begin{figure}[ht!]
	\centering
	\includegraphics[width=0.95\textwidth]{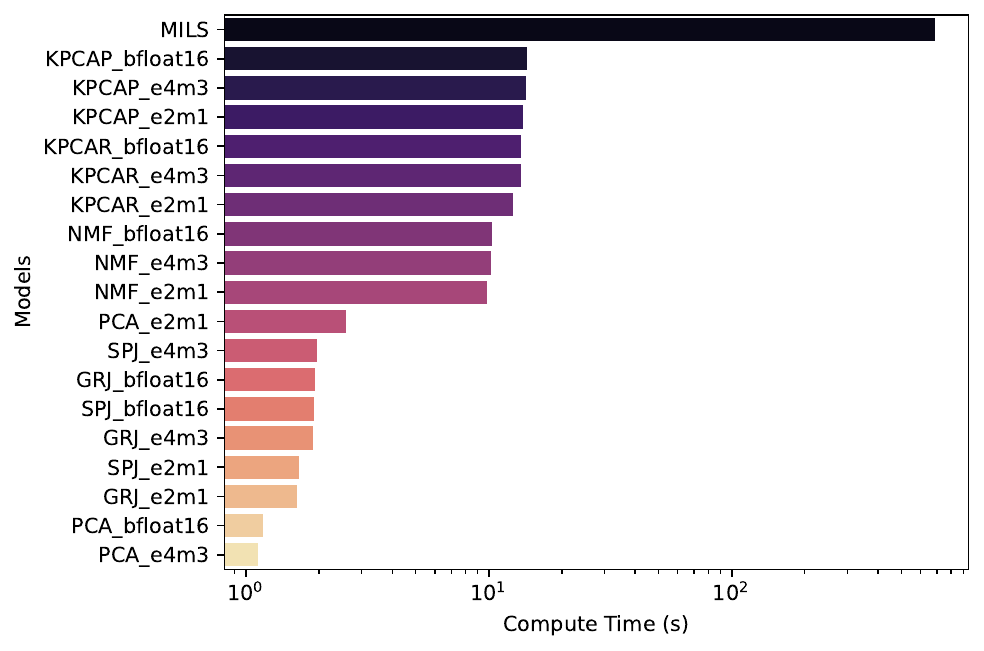}
	\caption{\label{bencht} Compute time in seconds for all the methods considered in this particular ML pipeline. MILS is less memory intensive at the cost of increasing computing times.}
\end{figure}

Figure \ref{bencht}, however, shows that the MILS implementation used leads to high computational cost (hereby expressed as compute time) when compared to other methods. The compute time for MILS is at least an order of magnitude higher than conventional methods, largely due to its reliance on pre-computed algorithmic complexity measures that are only available for small (4x4) patches. This limitation may be mitigated through the optimisation of MILS-specific implementation, including more efficient parallel computing strategies. Also, it is crucial to expand the precomputed Complexity-Theoretic Measure (CTM) database for image patches larger than 4×4, reducing lookup overhead.

Overall, MILS presents a highly promising alternative to traditional statistical Machine Learning dimensionality reduction algorithms, particularly in quantized ML scenarios. Its ability to leverage algorithmic complexity for superior information preservation makes it especially valuable for applications in Large Language Models (LLMs) and retrieval-augmented systems (RAGs) because MILS relies exclusively on binary operations, with potential for hardware acceleration and similarity search optimisation. Future work should focus on refining its computational efficiency to facilitate broader adoption in practical ML applications.

\subsection{Comparisons with other data dimensionality reduction methods}\label{sectionComparison}

\begin{figure}[ht!]
	\centering
	\scalebox{.3}{\includegraphics{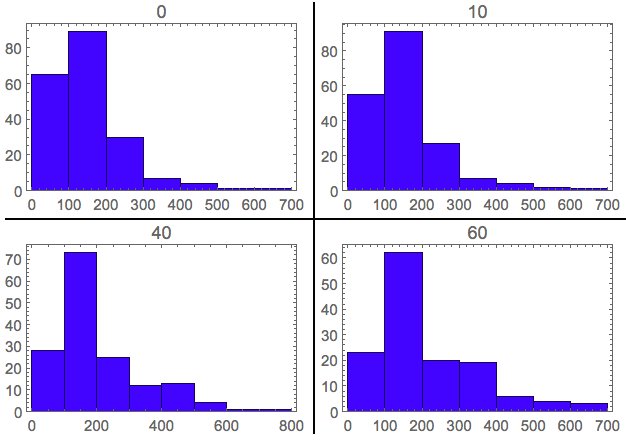}}
	\scalebox{.3}{\includegraphics{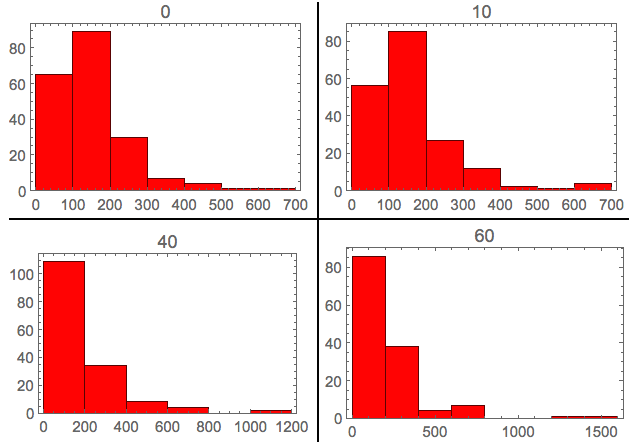}}\\
	\hspace{.1cm}
	\scalebox{.26}{\includegraphics{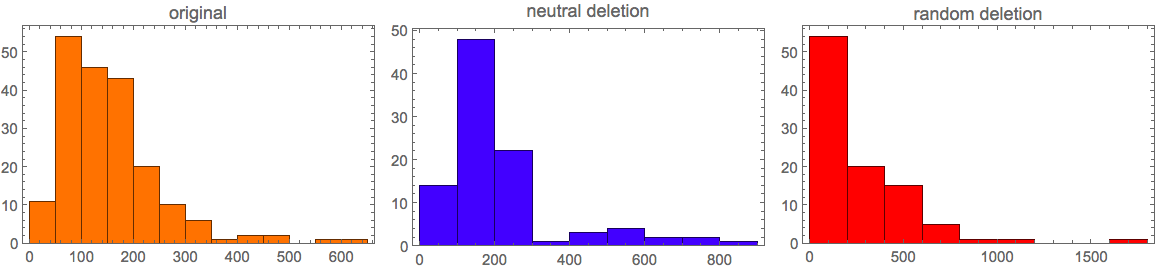}}
	\caption{\label{1}MILS or \emph{neutral edge} deletion (blue) outperforms random edge deletion (red) at preserving both edge degree distribution (top, showing removed edges) and edge betweenness distribution (bottom) on an Erd\H{o}s-R\'enyi random graph of vertex size 100 and low edge density ($\sim 4\%$) after up to 60 edges were removed (degree distribution comparison) and 150 edges were removed (edge betweenness) out of a total of 200 edges (notice also the scale differences on the $x$-axis).}
\end{figure}

\begin{figure}[ht!]
	\centering
	\scalebox{.25}{\includegraphics{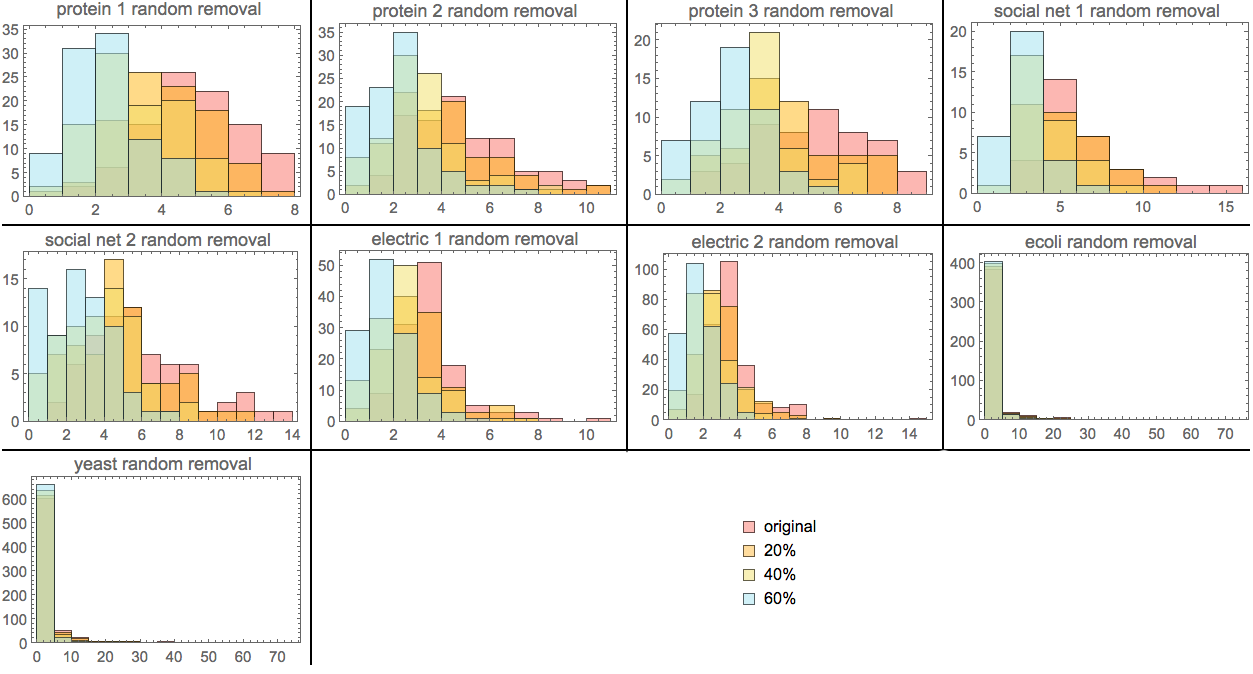}}\\
	\scalebox{.29}{\includegraphics{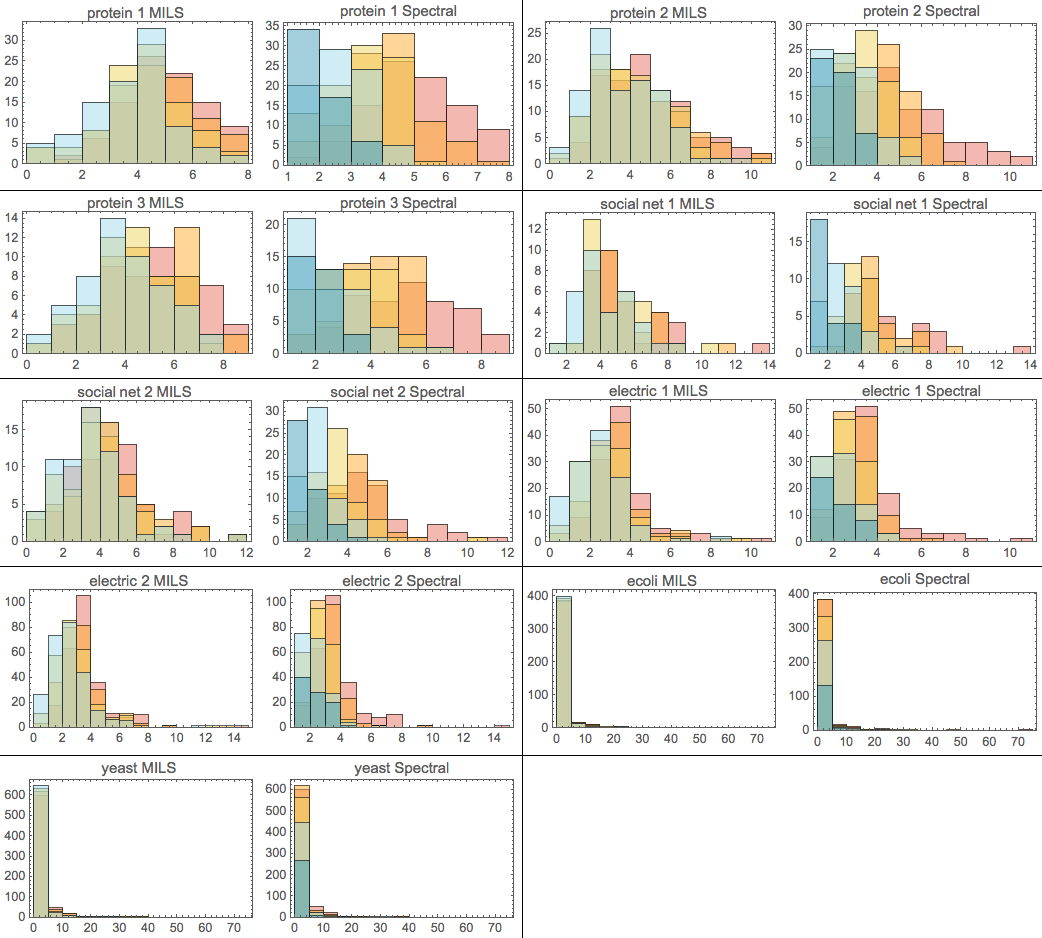}}
	\vspace{-0.3cm}\caption{\label{6}Histograms showing preservation of degree distribution from 20\% to 80\% edge removal. Green highlights the overlapping and the preserved area of the distributions after random deletion (top), MILS and spectral removal (bottom pairs).}
\end{figure}

\begin{figure}[ht!]
	\centering
	\scalebox{.35}{\includegraphics{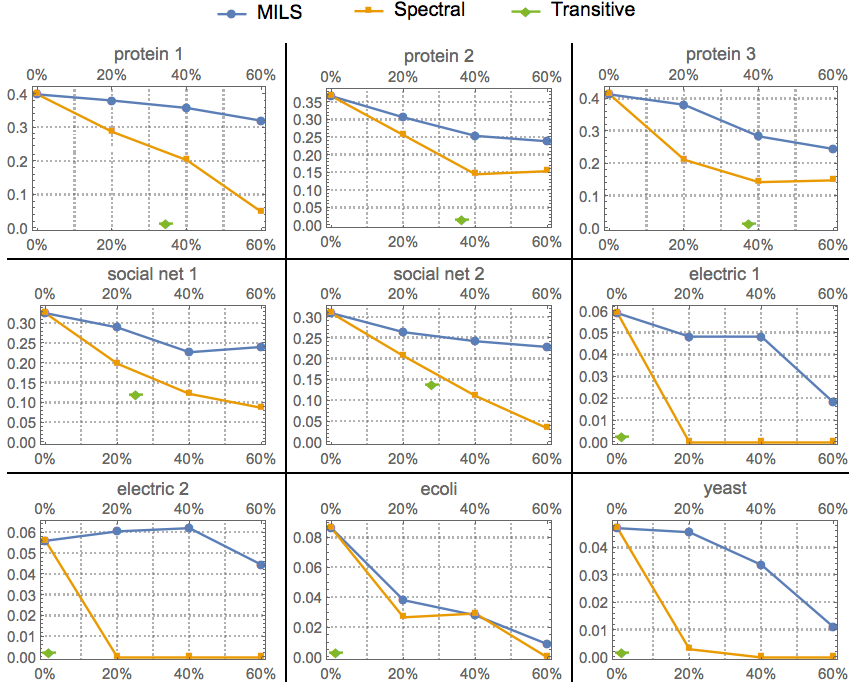}}
	\caption{\label{randomcluster}MILS mean clustering coefficient preservation against two sophisticated graph sparsification methods based on graph spectral and transitive reduction techniques on biological, electric and social networks taken from~\cite{superfamilies}. The transitive method does not allow selection of edges to be deleted, and in some cases it either fails to significantly reduce the network edge density if no cycles are present (such as, generally, in electric and genetic networks) and/or takes the clustering coefficient to 0 (e.g. for protein networks) if cycles are only local. Comparisons with other methods are unnecessary because they destroy local or global properties by design, such as clustering coefficients for the spanning tree algorithm.}
\end{figure}

\begin{figure}[ht!]
	\centering
	\scalebox{.36}{\includegraphics{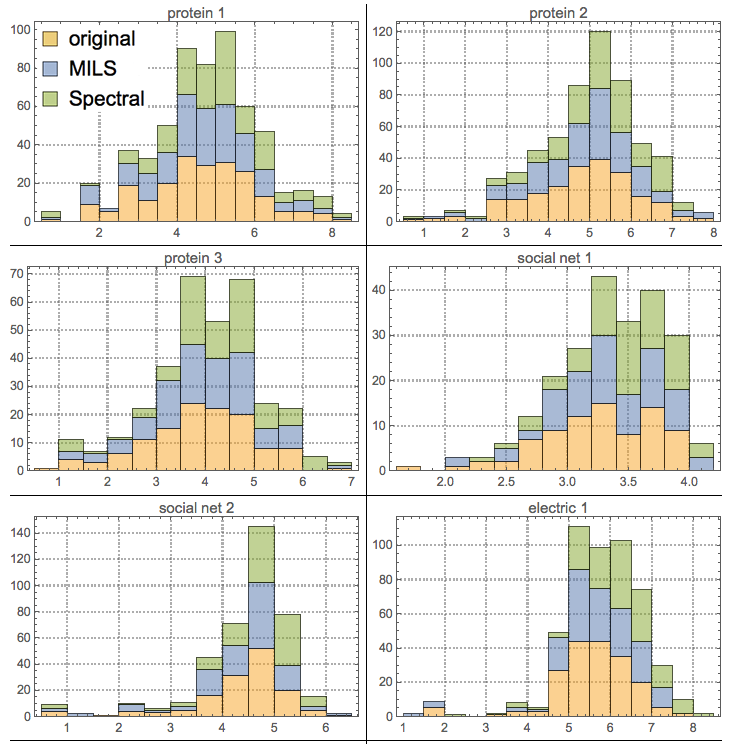}}\\
	\vspace{-.08cm}
	\hspace{.265cm}\scalebox{.365}{\includegraphics{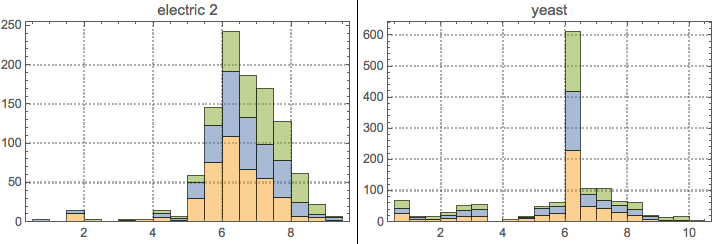}}
	\caption{\label{2}Stacked histograms showing edge betweenness preservation of MILS versus spectral sparsification across 
		different families of
		networks. The similarity in height of each segment is an indication of the preservation of such properties. Blue bars (MILS) approximate yellow (original) bars better than spectral sparsification. On average MILS was 1.5 times the edge betweenness distribution of these representative graphs measured by the area similarity of the respective bars.}
\end{figure}

\begin{figure}[ht!]
	\centering
	\scalebox{.37}{\includegraphics{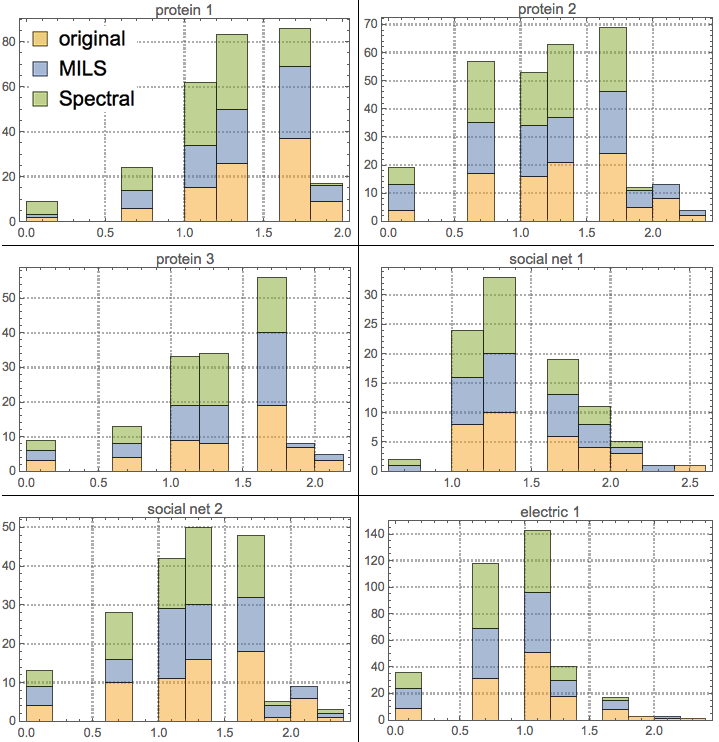}}\\
	\scalebox{.37}{\includegraphics{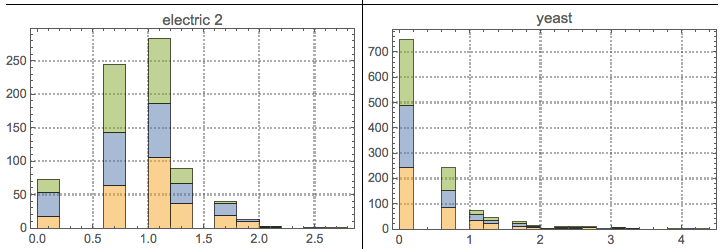}}
	\caption{\label{3}Stacked histograms showing the preservation of degree centrality after application of MILS versus spectral sparsification across different families of networks: bars with height closest to the original graph signify better preservation. Blue bars (MILS) approximate yellow (original) bars compared with spectral sparsification. MILS only slightly outperformed spectral sparsification in this test but never did worse.}
\end{figure}

\begin{figure}[ht!]
	\centering
	\scalebox{.37}{\includegraphics{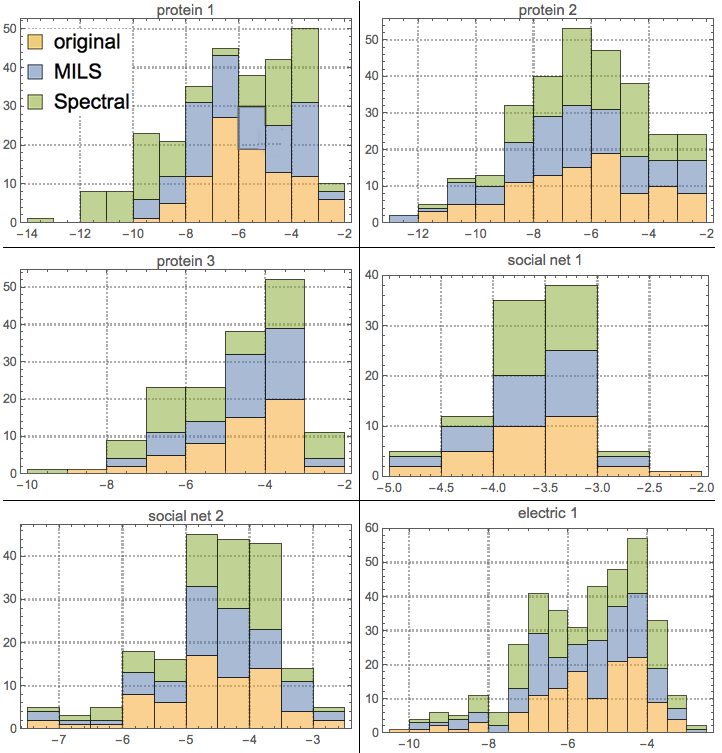}}\\
	\vspace{-.1cm}\hspace{.07cm}\scalebox{.37}{\includegraphics{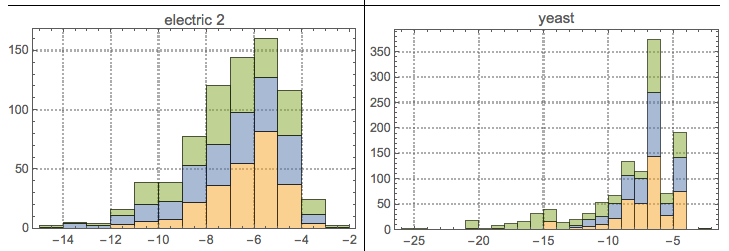}}
	\caption{\label{4}Stacked histograms showing eigenvector centrality preservation of MILS versus spectral sparsification across the different families of networks: bars with height closest to the graph's original bar signify better preservation. Blue bars (MILS) approximate yellow (original) bars better than spectral sparsification both in distribution shape and individual bar height. On average MILS preserved the eigenvector centrality distribution of these representative networks 1.5 times better.}
\end{figure}

Figs.~\ref{1},~\ref{6},~\ref{randomcluster},~\ref{2},~\ref{3}, and~\ref{4} 
illustrate the effectiveness of the Minimal Information Loss Selection (MILS) method in retaining crucial local and global characteristics of synthetic and natural networks, encompassing varied types and topologies. These figures demonstrate MILS's performance, typically surpassing leading graph sparsification algorithms.
The study involved well-known networks from~\cite{superfamilies}, such as genetic regulatory, protein, power grid, and social networks. MILS was applied to these networks for comparative analysis with transitive reduction~\cite{aho} and spectral sparsification~\cite{spielman}, two powerful sparsification methods. The former seeks to minimise edges while maintaining the same reachability within a directed graph, while the latter, thoroughly introduced in~\cite{batson}, aims to reduce network dimensions based on spectral similarity of graph Laplacians, thus preserving significant properties through the adjacency matrix Laplacian spectrum.

Specifically, Figs.~\ref{1} and~\ref{6} showcase the performance of MILS in preserving the degree distribution and edge betweenness distribution in a typical synthetically (recursively) generated Erd\H{o}s-R\'{e}nyi (ER) random graph with low edge density, comparing its results against both random edge deletion and spectral sparsification methods.



While MILS is not significantly better at preserving the clustering coefficient of random networks, Fig.~\ref{randomcluster} shows that MILS does significantly better at preserving the clustering coefficient of real-world (biological, social, and electric grid) networks taken from~\cite{superfamilies}, outperforming both transitive and spectral reduction/sparsification methods. 

Figs.~\ref{2}, \ref{3} and \ref{4} illustrate how MILS outperforms spectral sparsification at preserving edge betweenness, and degree and eigenvector centralities, respectively.

\subsection{An application of MILS to image lossy compression}\label{sectionMILSonimages}

\begin{figure}[ht!]
	\centering
	\scalebox{.43}{\includegraphics{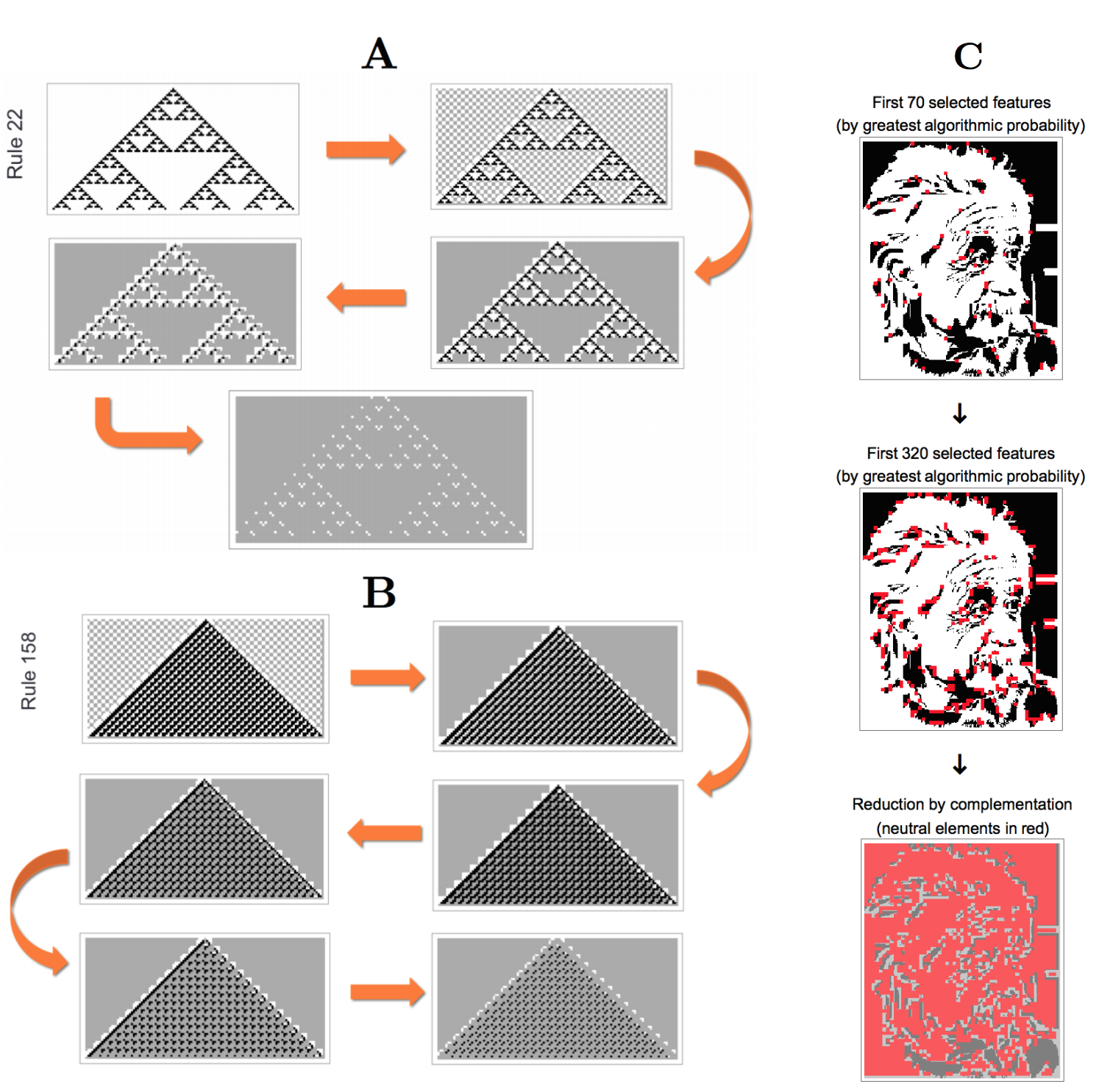}}
	
	\caption{\label{a}In (A) and (B) we show a feature selection and image reduction by application of MILS, starting from the original and second step, highlighting the regions that are earmarked to be omitted (in grey) versus the features that are kept along the way, thereby optimally preserving the main properties of these objects, properties whose persistence enables a ranking of such features. Here can be seen how boundaries are favoured as key features. In (C) we see the image reconstruction by preservation of extracted features from the image with the highest algorithmic probability (indicated by red signatures in the first two image processes). The MILS algorithm selects features which are considered most important in the compressed image reconstruction.}
\end{figure}

Fig. \ref{a}(A) and Fig. \ref{a}(B) 
illustrate the application of MILS in the reduction and coarse-graining of two simple cases of the space-time evolution of Elementary Cellular Automata (ECA) rules 22 and 158. The reduction is by minimisation of algorithmic information loss. MILS effectively extracts the salient elements that characterise each of these systems. 
The examples in Fig. \ref{a}(A) and Fig. \ref{a}(B) depicts the way in which regions with high or low algorithmic content can be ranked, selected, or preserved for dimensional reduction purposes. The method performs an unsupervised lossy compression able to preserve both cellular automata's main features (not covered in grey), with no intervention and no parameter choice. MILS proceeds by deleting the regions with the lowest algorithmic information content contribution and maximizing the preservation of the features that contribute the most to the algorithmic description of the objects.
The extracted features will not be as clear as in these examples as they may pick more complicated patterns, even not statistical ones based on algorithmic probability. Unlike statistical approaches, the algorithm can approximate (and thus preserve/extract) features that are of an algorithmic nature and which are not statistically apparent as it was in this case (see~\cite{bdm,zkgraph}) and next examples.

%

\begin{figure}[ht!]
	\centering
	\scalebox{.5}{\includegraphics{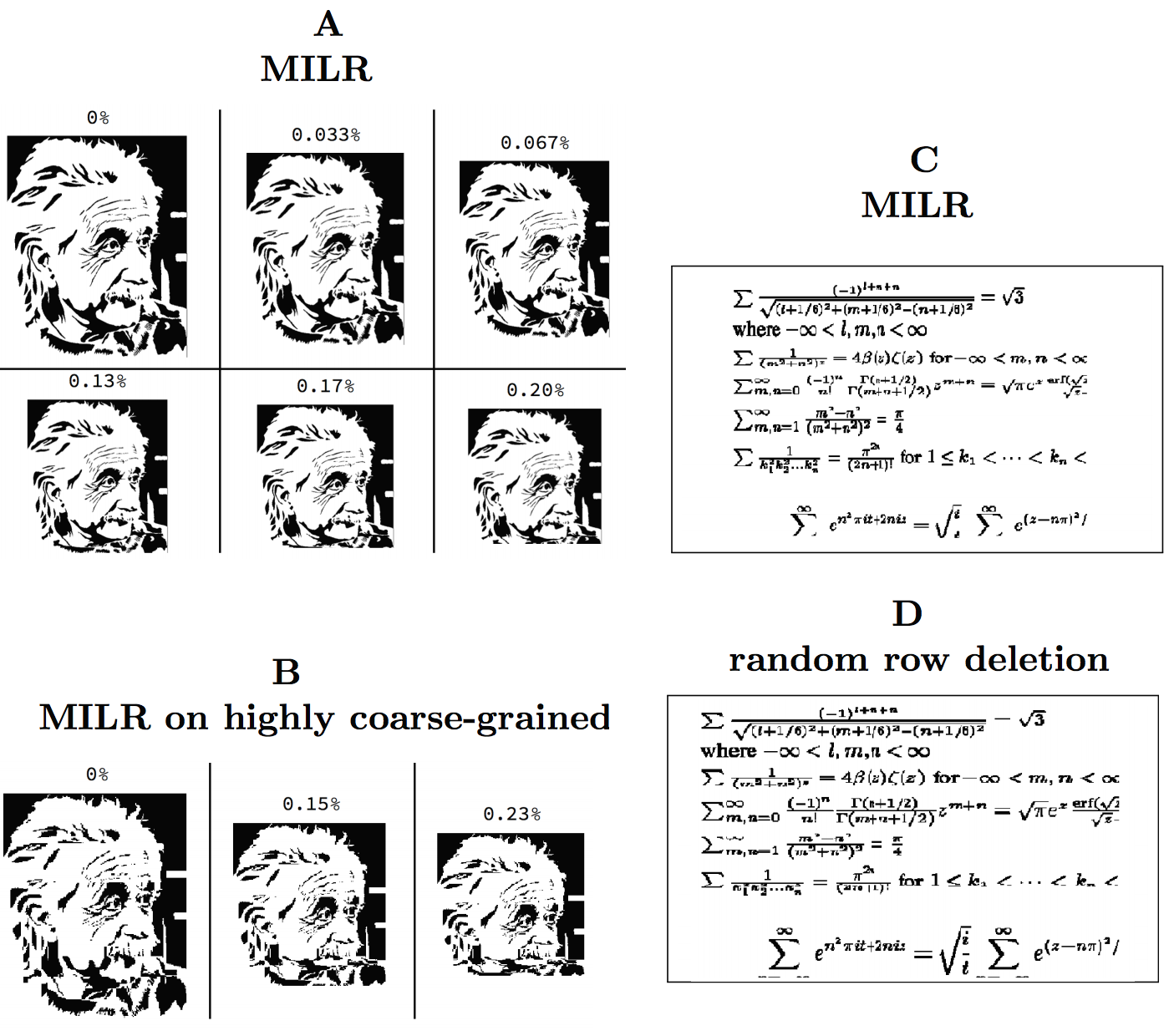}}
	\caption{\label{b}Minimal Information Loss Reduction, a MILS-based lossy compression algorithm. A: Row compression preserves features on even highly coarse-grained versions (B starts from A at 0\% but sampled to make it a 100 $\times$ 100 image from the original 600 $\times$ 600) of the same image producing a cartoonish representation. The image in (B) reduced by 23\% is 10\% the size of the image at 0\% in (A). C: Compressed formula by MILS vs random row deletion (D) with resulting images of the same size but MILS preserving text proportion and minimizing information loss while random row deletion distorts the text.}
\end{figure}

Fig.~\ref{b} shows the MILS algorithm applied to images. 
Fig.~\ref{b}(A) shows how vertical and horizontal compression preserves features even if distorting the image hence showing different mechanisms and goals than those from popular image compression algorithms. The purpose of image compression algorithms such as JPEG is to maximise storage compressibility and require a decoder, here the application is directly to the image itself both at storage and visualisation stages. Is also important to mention that our purpose is not to maximise storage compression but to minimise the loss of algorithmic content in the reduction process. 
Fig.~\ref{b}(C) shows how the algorithm preserves the main features of the image leaving almost intact the formulae.

MILS is highly context-sensitive and adapts to different kind of information such as images. 
For example, it handles in different ways images as seen in Fig.~\ref{b}.
This is because information in Fig.~\ref{b}(A) is distributed more uniformly than Fig.~\ref{b}(C) but, according to the algorithm, some features in certain regions are more important than others and thus some distortions are allowed if such features have high algorithmic content according to the underlying computer programs that reconstruct the compressed images. 
However, in Fig. \ref{b}(C), blank space between formulae can be sacrificed first as they do not contain any information. While this behaviour for handling different cases can be replicated in ad-hoc algorithms, MILS handles it naturally and determines when distorting can be allowed or proportion can be preserved, something that image compression algorithms would not do as their goal is to recover the same image dimensions and maximise storage compression and not to preserve algorithmic information content even sacrificing otherwise fundamental properties of the image for those algorithms such as image dimension.

\section{Conclusion}\label{Conclusion}

We have introduced a set of dimensionality reduction techniques based on a Minimal Information Loss Selection (MILS) algorithm, a recursive unsupervised Machine Learning approach particularly effective in reducing data dimensionality preserving essential features in data. In the context of networks, we have shown it preserves features such as degree distribution, clustering coefficient, edge betweenness, and centralities better than traditional and state-of-the-art algorithms. As studied in Section~\ref{sectionComparison},  MILS demonstrates comparable or superior performance to existing network reduction algorithms on different classes and types of networks. Moreover, it extends to lossy compression of images and bi-dimensional data as shown in Section~\ref{sectionMILSonimages}. It operates on the principle of local perturbation analysis and Algorithmic Information Dynamics, offering robustness against varying methods of approximating algorithmic complexity.
The method presented in this article optimally preserves the algorithmic information content (which might be affected by the local perturbations) whenever an optimal method is used to approximate algorithmic complexity, effectively introducing a lossy technique based on principles of lossless recursive compression.

Its reliance on first principles guarantees that its performance can only increase as more computational resources are made available, which differs from other methods based on traditional statistical Machine Learning for which theoretical limitations are known to hold~\cite{zkgraph}. Nevertheless, the empirical application of MILS only scales linearly with the worst-case time complexity of the chosen method for approximating algorithmic complexity.

\section*{Funding}
FSA acknowledges support from the São Paulo Research Foundation (FAPESP), grants $2023$/$05593$-$1$ and $2021$/$14501$-$8$. NAK is supported by a Horizon $2020$ EU grant (MultipleMS, $733161$).

\newpage
\appendix


\begin{thebibliography}{100}
	
	
	
	
	

	\bibitem{aho} A.V. Aho, M.R. Garey, J.D.Ullman, The transitive reduction of a directed graph, \textit{SIAM Journal on Computing} 1 (2): 131--137, 1972.
    
    	\bibitem{batson} J. Batson, D.A. Spielman, N. Srivastava, and S.-H. Teng, Spectral Sparsification of Graphs: Theory and Algorithms, vol. 56:8, \textit{Communications of the ACM}, 2013.

	\bibitem{boehmke} B. Boehmke, B.M. Greenwell, Dimension Reduction, \textit{Hands-On Machine Learning with R}, Chapman \& Hall. pp. 343--396, 2019.

	\bibitem{chew} P. Chew. There are planar graphs almost as good as the complete graph, \textit{J. Comput. Syst. Sci.,} 39:205--219, 1989.

	\bibitem{cunningham} P. Cunningham, Dimension Reduction (Technical report). University College Dublin. UCD-CSI-2007-7, 2007.
    
	\bibitem{fodor} I. Fodor, A survey of dimension reduction techniques (Technical report). Center for Applied Scientific Computing, Lawrence Livermore National. UCRL-ID-148494, 2002.

	\bibitem{Liu2018}
	Y. Liu, T. Safavi, A. Dighe, and D. Koutra.
	\newblock {Graph Summarization Methods and Applications}.
	\newblock {\em ACM Computing Surveys}, 51(3):1--34, jun 2018.

	\bibitem{bencz} A. Benczur and D.R. Karger. Approximating s-t minimum cuts in O(n2)time. \textit{In Proceedings of The Twenty-Eighth Annual ACM Symposium On The Theory Of Computing (STOC 96)}, pages 47--55, May 1996.    

	\bibitem{spielman} D.A. Spielman, N. Srivastava, Graph sparsification by effective resistances, \textit{Proceedings of the fortieth annual ACM symposium on Theory of computing (STOC '08),} 563--568, 2008.

	\bibitem{zenildata} H. Zenil, Algorithmic Data Analytics, Small Data Matters and Correlation versus Causation. In M. Ott, W. Pietsch, J. Wernecke (eds.), \textit{Berechenbarkeit der Welt? Philosophie und Wissenschaft im Zeitalter von Big Data}, Springer Verlag, pp. 453--475, 2017.

	\bibitem{zkgraph} H. Zenil, N.A. Kiani and J. Tegn\'er, Low Algorithmic Complexity Entropy-deceiving Graphs, \textit{Physics Reviews E.} 96, 012308, 2017.
	
	\bibitem{graphreview} H. Zenil, N.A. Kiani, J. Tegn\'er, A Review of Graph and Network Complexity from an Algorithmic Information Perspective, \textit{Entropy}, 20(8):551, 2018.



	\bibitem{Kiani2016InferenceGeneticnetworks} N.~A. Kiani, H. Zenil, J. Olczak, and J. Tegn{\'{e}}r.
	\newblock {Evaluating network inference methods in terms of their ability to preserve the topology and complexity of genetic networks}.
	\newblock {\em Seminars in Cell \& Developmental Biology}, 51:44--52, 2016.	 


	\bibitem{postchina} H. Zenil, N.A. Kiani and J. Tegn\'er, Quantifying Loss of Information in Network-based Dimensionality Reduction Techniques, \textit{Journal of Complex Networks}, vol. 4:(3) pp. 342-362, 2016.

	
	\bibitem{numerical} F. Soler-Toscano, H. Zenil, J.-P. Delahaye and N. Gauvrit, Correspondence and Independence of Numerical Evaluations of Algorithmic Information Measures, \textit{Computability}, vol. 2, no. 2, pp. 125--140, 2013.

	\bibitem{bdm} H. Zenil, F. Soler-Toscano, N.A. Kiani, S. Hern\'andez-Orozco, A. Rueda-Toicen, A Decomposition Method for Global Evaluation of Shannon Entropy and Local Estimations of Algorithmic Complexity, \textit{Entropy} 20(8), 605, 2018. 

	\bibitem{Calude2002} C.S. Calude.
	\newblock {\em {Information and Randomness: An algorithmic perspective}}.
	\newblock Springer-Verlag, 2 edition, 2002.

	\bibitem{d4} J.-P. Delahaye and H. Zenil, Numerical Evaluation of the Complexity of Short Strings: A Glance Into the Innermost Structure of Algorithmic Randomness, \textit{Applied Mathematics and Computation} 219,  63--77, 2012.

	\bibitem{d5} F. Soler-Toscano, H. Zenil, J.-P. Delahaye and N. Gauvrit, \textit{Calculating Kolmogorov Complexity from the Frequency Output Distributions of Small Turing Machines}, PLoS ONE 9(5), e96223, 2014.

	\bibitem{kolmo2d} H. Zenil, F. Soler-Toscano, J.-P. Delahaye and N. Gauvrit, Two-Dimensional Kolmogorov Complexity and Validation of the Coding Theorem Method by Compressibility, \textit{PeerJ Computer Science}, 1:e23, 2015.

	\bibitem{Abrahao2021} F. S. Abrahão, K. Wehmuth, H. Zenil, and A. Ziviani, “An Algorithmic Information Distortion in Multidimensional Networks,” in \textit{Complex Networks and Their Applications IX}, Cham, 2021, vol. 944, pp. 520–531. doi: 10.1007/978-3-030-65351-4{\_}42.

	\bibitem{Abrahao2021a} F. S. Abrahão, K. Wehmuth, H. Zenil, and A. Ziviani, “Algorithmic Information Distortions in Node-Aligned and Node-Unaligned Multidimensional Networks,” \textit{Entropy}, vol. 23, no. 7, p. 835, Jun. 2021, doi: 10.3390/e23070835.
	
	\bibitem{Abrahao2021bpublished}
	F.~S. Abrahão, H.~Zenil, Emergence and algorithmic information dynamics of
	systems and observers, Philosophical Transactions of the Royal Society A:
	Mathematical, Physical and Engineering Sciences 380~(2227) (2022).
	\newblock \href {https://doi.org/10.1098/rsta.2020.0429}
	{\path{doi:10.1098/rsta.2020.0429}}.

	\bibitem{maininfo} H. Zenil, N.~A. Kiani, F. Marabita, Y. Deng, S. Elias,
	A. Schmidt, G. Ball, and J. Tegn{\'{e}}r.
	\newblock {An Algorithmic Information Calculus for Causal Discovery and
		Reprogramming Systems}.
	\newblock {\em iScience}, 19:1160--1172, sep 2019.

	\bibitem{Zenil2020}
	H.~Zenil, A review of methods for estimating algorithmic complexity: options, challenges, and new directions, Entropy 22~(6) (2020) 612.
	\newblock \href {https://doi.org/10.3390/e22060612}
	{\path{doi:10.3390/e22060612}}.

	\bibitem{Hernandez-Orozco2018a}
	S. Hern{\'{a}}ndez-Orozco, N.~A. Kiani, and H. Zenil.
	\newblock {Algorithmically probable mutations reproduce aspects of evolution,
		such as convergence rate, genetic memory and modularity}.
	\newblock {\em Royal Society Open Science}.
	
	\bibitem{HernandezOrozco2021AlgProbML}
	Santiago Hernández-Orozco, Hector Zenil, Jürgen Riedel, Adam Uccello,
	  Narsis~A. Kiani, and Jesper Tegnér.
	\newblock Algorithmic {Probability}-{Guided} {Machine} {Learning} on
	  {Non}-{Differentiable} {Spaces}.
	\newblock {\em Frontiers in Artificial Intelligence}, 3:567356, January 2021.

	\bibitem{Zenil2019CausalDeconv}
	Hector Zenil, Narsis~A. Kiani, Allan~A. Zea, and Jesper Tegn{\'{e}}r.
	\newblock {Causal deconvolution by algorithmic generative models}.
	\newblock {\em Nature Machine Intelligence}, 1(1):58--66, jan 2019.
        \bibitem{Uthamacumaran2023arxivMSpaper}Uthamacumaran, A., Abrahão, F., Kiani, N. \& Zenil, H. On the salient limitations of the methods of assembly theory and their classification of molecular biosignatures. {\em Npj Systems Biology And Applications}. \textbf{10} (2024,8), http://dx.doi.org/10.1038/s41540-024-00403-y
        \bibitem{Zenil2018bReprogrammabilityChemicalNetworks} H. Zenil, N.~A. Kiani, M. Shang, and J. Tegn{\'{e}}r.
	\newblock {Algorithmic {Complexity} and {Reprogrammability} of {Chemical} {Structure} {Networks}}.
	\newblock {\em Parallel Processing Letters}, vol. 28:(1), 2018.

	\bibitem{Zenil2020cnat}
	H.~Zenil, N.~Kiani, F.~Abrah{\~{a}}o, J.~Tegn{\'{e}}r, {Algorithmic Information
		Dynamics}, Scholarpedia Journal 15~(7) (2020) 53143.
	\newblock \href {https://doi.org/10.4249/scholarpedia.53143}
	{\path{doi:10.4249/scholarpedia.53143}}.

	\bibitem[Zenil et~al.(2023{\natexlab{b}})Zenil, Kiani, and Tegner]{algodyn2}
	Hector Zenil, Narsis Kiani, and Jesper Tegner.
	\newblock \emph{{Algorithmic Information Dynamics}}.
	\newblock Cambridge University Press, 1 edition, 2023{\natexlab{b}}.
	\newblock ISBN 1108497667.

	\bibitem{zenilkianitegner} H. Zenil, N.A. Kiani and J. Tegn\'er, Methods of Information Theory and Algorithmic Complexity for Network Biology, \textit{Seminars in Cell and Developmental Biology}, vol. 51, pp. 32-43, 2016.

	\bibitem{Zenil2019b}
	H. Zenil, N.~A. Kiani, and J. Tegn{\'{e}}r.
	\newblock {The Thermodynamics of Network Coding, and an Algorithmic Refinement of the Principle of Maximum Entropy}.
	\newblock {\em Entropy}, 21(6):560, jun 2019.
	
	\bibitem{Downey2010}
	R.G. Downey and D.R. Hirschfeldt.
	\newblock {\em {Algorithmic Randomness and Complexity}}.
	\newblock Theory and Applications of Computability. Springer New York, New
	York, NY, 2010.

    \bibitem{vi1} M. Li and P.M.B. Vit\'anyi, An \textit{Introduction to Kolmogorov Complexity and Its Applications}, 3rd. Ed. Springer, 2008.
	
	\bibitem{algodyn} H. Zenil, N.A. Kiani and J. Tegn\'er, Algorithmic Information Dynamics of Emergent, Persistent, and Colliding Particles in the Game of Life. In A. Adamatzky (ed), \textit{From Parallel to Emergent Computing}, Taylor \& Francis / CRC Press, pp.367--383, 2019.

	
	\bibitem{zenilgraph} H. Zenil, F. Soler-Toscano, K. Dingle, and A.~A. Louis.
	\newblock {Correlation of automorphism group size and topological properties
		with program-size complexity evaluations of graphs and complex networks}.
	\newblock {\em Physica A: Statistical Mechanics and its Applications},
	404:341--358, 2014.

    \bibitem{Calude2022GlimpseOmega} Cristian~S. Calude, Michael~J. Dinneen, and Chi-Kou Shu.
	\newblock Computing a glimpse of randomness.
	\newblock {\em Experimental Mathematics}, 11(3):361--370, 2002.

    \bibitem{Bengio}Bengio, Y., Delalleau, O., Roux, N., Paiement, J., Vincent, P. \& Ouimet, M. Learning Eigenfunctions Links Spectral Embedding and Kernel PCA. {\em Neural Computation}. \textbf{16}, 2197-2219 (2004)

    \bibitem{superfamilies} R. Milo, S. Itzkovitz, N. Kashtan, R. Levitt, S. Shen-Orr, V. Ayzenshtat, M. Sheffer, U. Alon, Superfamilies of designed and evolved networks, \textit{Science} 303, 1538--1542, 2004.

	
	
	
	
	
	
	
	
	
	

	
	
	
	
	
	
	
	
	
	
	
	
	
	
	
	
	
	
	

	
	

	
	
	
	
	
	
	
	
	
	
	

	

	
	
	
	
	
	
\end{thebibliography}
\end{document}